\documentclass[12pt, draftclsnofoot, onecolumn]{IEEEtran}
\usepackage{cite}
\usepackage{amsmath,amssymb,amsfonts}
\usepackage{algorithmic}
\usepackage{graphicx}
\usepackage{textcomp}
\usepackage{amsmath}
\usepackage{amssymb}
\usepackage{latexsym}
\usepackage{CJK}
\usepackage{booktabs}
\usepackage{array}
\usepackage{multirow}
\usepackage{graphicx}
\usepackage{epstopdf}
\usepackage{caption}
\usepackage{makecell}
\usepackage{threeparttable}
\usepackage{bm}
\usepackage{subfigure}

\usepackage{algorithm}
\usepackage{algorithmic}
\usepackage{tabularx}
\usepackage{stfloats}
\usepackage{xcolor}
\usepackage{booktabs}

\ifCLASSINFOpdf
\else
\fi

\begin{document}
%
\title{Joint Hardware Design and Capacity Analysis for Intelligent Reflecting Surface Enabled Terahertz MIMO Communications}


\author{

      Xinying Ma, Zhi Chen, Chongwen Huang, Longfei Yan, \\Chong Han, and Qiye Wen
\thanks{

X. Ma, Z. Chen, W. Chen, and Q. Wen are with University of Electronic Science and Technology of China, Chengdu, China (e-mails: xymarger@126.com; chenzhi@uestc.edu.cn; qywen@uestc.edu.cn). C. Huang is with Zhejiang University, Hangzhou, China (e-mails: chongwenhuang@zju.edu.cn). L. Yan and C. Han are with Shanghai Jiao Tong University, Shanghai, China (e-mails: longfei.yan@sjtu.edu.cn; chong.han@sjtu.edu.cn).}
}


\maketitle

\begin{abstract}
Terahertz (THz) communications have been envisioned as a promising enabler to provide ultra-high data transmission for sixth generation (6G) wireless networks. To tackle the blockage vulnerability brought by severe attenuation and poor diffraction of THz waves, a reconfigurable intelligent surface (RIS) is put forward to smartly manipulate the propagation directions of incident THz waves. In this paper, we firstly design an efficient and practical hardware structure of graphene-based RIS that phase response can be programmed up to 306.82 degrees. Subsequently, to achieve the hybrid beamforming design of the RIS-enabled THz multiple-input multiple-output (MIMO) system, an adaptive gradient descent (A-GD) algorithm is developed by dynamically updating the step size during the iterative process, which is determined by the second-order Taylor expansion formulation. In contrast with the conventional gradient descent (C-GD) algorithm with fixed step size, the A-GD algorithm improves the achievable rate performance significantly. However, both A-GD and C-GD algorithms have the inherited heavy complexity. Then a low complexity alternating optimization algorithm is proposed by alternately optimizing the precoding matrix via a column-by-column algorithm and the phase shift matrix of the RIS via a linear search algorithm. Finally, numerical results demonstrate the effectiveness of our designed hardware structure and developed algorithms.
\end{abstract}



\section{Introduction}
With the continuous explosion growth of data traffic in wireless communications, sixth generation (6G) communication networks are expected to meet a great deal of pressing requirements in the near future, such as increased spectral efficiency, higher data rate, lower latency, larger connection density and so on \cite{introduction_01,introduction_02}. To settle these challenges, Terahertz (THz) frequency band (0.1-10 THz) has been regarded as a prospective alternative to provide large spectrum bandwidth and support ultra-high data transmission for 6G communication networks \cite{introduction_03}. Since THz communication is able to realize high transmission rates from hundreds of gigabits per second (Gbps) to several terabits per second (Tbps), some typical application scenarios are defined and considered recently, including intra-device communications, high speed kiosk downloads, wireless data centers and wireless backhaul networks\cite{introduction_04}. From the perspective of spectrum resources, THz frequency band bridges the gap between millimeter wave (mmWave) and optical frequency ranges \cite{introduction_06}. Compared with mmWave frequency band, THz communication possesses much larger bandwidth and better security performance. In contrast with optical frequency band, THz communication is much easier to realize the beam tracking, and adapts inconvenient climate conditions. Enabled by these obvious advantages, THz communication is regarded as an indispensable technology for 6G communication networks.

Despite the numerous advantages, there are still some imperative challenges existing in THz communication systems. On the one hand, due to the high path attenuation and strong molecular absorption effect experienced by THz waves, the transmission distance of THz communications is limited within a small area, and thus is applicable for the specific communication scenarios \cite{introduction_08}, \cite{introduction_09}. On the other hand, THz waves at such a high frequency band undergo extremely poor diffraction, and THz communication links that depend on the line-of-sight (LoS) path are easily blocked by the obstacles. Given this, the concept of a reconfigurable intelligent surface (RIS) is newly proposed to mitigate blockage vulnerability and improve coverage capability \cite{yanlongfei_01,introduction_10,introduction_10-1,introduction_12,introduction_13}. To be specific, the RIS is a kind of physical meta-surface consisting of a large number of passive reflecting elements. Each reflecting element is capable of adjusting the phase shifts by using a smart processor \cite{introduction_13-1}. Similar to backscatter communications \cite{introduction_14} and smart reflect-array \cite{introduction_16}, the reflecting elements of RIS are passive and lack the active radio frequency (RF) chains, and thus is more energy-efficient compared with existing solutions based on active RF devices, such as amplify-and-forward relaying \cite{introduction_17} and massive MIMO. Therefore, the RIS-enabled THz communication systems are worthy of further exploration.

Deploying RISs in the THz communication system is essential, but some challenges also emerge accordingly. To realize reliable THz communications, the channel state information (CSI) acquisition is the primary mission before the data transmission begins. Different from conventional communication systems with active devices, the main difficulty of channel estimation problem in RIS-enabled THz systems is that the reflecting elements are passive and are unable to achieve the signal processing. Prominently, by leveraging the sparse features of THz multiple-input multiple-output (MIMO) channel, the authors convert the channel estimation problem into the sparse signal recovery problem, and a low complexity compressed sensing (CS) based channel estimation scheme is developed to realize the efficient signal reconstruction \cite{introduction_21}. In addition, a two-stage channel estimation algorithm that includes a sparse matrix factorization stage and a matrix completion stage is developed in \cite{introduction_21-1}, and a novel message-passing based channel estimation algorithm is proposed to solve the matrix-calibration based matrix factorization problem in \cite{introduction_21-2}. Once the CSI is acquired at the base station (BS) side or the mobile station (MS) side, the extensive research directions can be investigated, such as energy efficiency optimization \cite{introduction_13, introduction_21-3}, data rate maximization \cite{introduction_22, introduction_23, introduction_24}, secure communication \cite{introduction_29}. There are also some RIS prototypes for THz communications \cite{introduction_29-1,introduction_29-2}, but the algorithm design are not actually taken into consideration. Apart from these aforementioned research interests, the joint hybrid beamforming and hardware design for the RIS-empowered THz MIMO communication system is still treated as an open problem.

In order to compensate for the research gap, the joint hardware design and hybrid beamforming optimization for the RIS-enabled THz MIMO system is presented in this paper, \emph{ which is the first attempt to practically combine the hardware characteristics of the RIS and the software design together}. Firstly, we develop a downlink RIS-enabled THz MIMO communication system model, and an efficient graphene-based hardware structure of each RIS element with a wide phase response range and a desired reflecting amplitude is designed. Then, considering the constraints of the hardware features, the gradient descent based algorithm and the alternating optimization (AO) algorithm are put forward to jointly optimize the phase shift matrix at the RIS and the hybrid beamforming matrix at the BS, which greatly differs from the previous works in \cite{introduction_31,introduction_31-1,introduction_31-2} that only consider the phase shift design of the RIS. Compared with the conventional MIMO system without the RIS \cite{introduction_32}, \cite{introduction_33}, the maximum rate optimization problem of the RIS-enabled THz MIMO system involves multiple matrix variables, and thus is more sophisticated. The main contributions of this paper can be summarized as follows.

\begin{itemize}
  \item To begin with, we design a practical graphene-based RIS hardware structure, where its  phase response can be controlled  up to 306.82 degrees, and the reflecting amplitude efficiency is more than 50{\%} at 1.6 THz. Furthermore, the design theory and working principle of the RIS are  also provided. Then, the electric properties of the graphene are introduced by revealing the relationship between conductivity and applied voltage, which is the foundation of forming an electrically-controlled RIS.

  \item Next, the adaptive gradient descent (A-GD) algorithm is developed to seek a high-quality hybrid beamforming design by maximizing the achievable rate for the RIS-enabled THz MIMO system. Specifically, compared with conventional gradient descent (C-GD) algorithm with the fixed step size, the proposed A-GD algorithm turns out to be more efficient by dynamically updating the step size during the iterative process, and is able to realize a better achievable rate performance.

  \item Moreover, to combat the heavy complexity brought by gradient descent based algorithms, a low complexity AO algorithm is raised by alternately optimizing the precoding matrix and the phase shift matrix of the RIS. On the one hand, a column-by-column (CBC) algorithm is proposed to optimize the precoding matrix. On the other hand, a linear search algorithm is considered to determine the phase shift matrix of the RIS by utilizing the one-the-rest criterion.

  \item Finally, simulation results demonstrate that our designed RIS structure with the practical phase response of 306.82 degrees achieves basically identical performance in contrast with the ideal phase response of 360 degrees. Specifically, the proposed A-GD algorithm exhibits the best achievable rate performance, while the developed AO algorithm achieves a better balance between the computational complexity and achievable rate performance.
\end{itemize}

The reminder of this paper is organized as follows. Section II describes the RIS-enabled THz MIMO system model and the channel model. In Section III, an effective hardware structure of graphene-based RIS is designed. Section IV discusses the proposed A-GD algorithm. Then, an AO algorithm is proposed by alternately operating the CBC scheme and the linear search scheme in Section V. The simulation results and conclusion are presented in Section VI and Section VII, respectively.

\emph{Notations:} ${{\bf{A}}}$ is a matrix, ${{\bf{a}}}$ is a vector, ${a}$ is a scalar. ${{\left\| {\bf{A}} \right\|_F}}$ is the Frobenius norm, whereas ${{{\bf{A}}^H}}$, ${{\bf{A}}^ * }$, ${{{\bf{A}}^T}}$, ${{{\bf{A}}^{ - 1}}}$, ${{{\bf{A}}^\dag }}$, ${\left| {\bf{A}} \right|}$ and $rank\left( {\bf{A}} \right)$ are conjugate transpose, conjugate, transpose, inverse, pseudo-inverse, determinant and the rank of ${\bf{A}}$, respectively. ${\mathop{\rm diag}\nolimits} \left( {\bf{a}} \right)$ is a diagonal matrix with elements of ${\bf{a}}$ on its diagonal. ${\mathop{\rm Tr}\nolimits} \left( {\bf{A}} \right)$ is the trace of matrix ${\bf{A}}$. ${{\mathop{\rm E}\nolimits} \left[  \cdot  \right]}$ is used to denote the expectation. ${\mathop{\rm vec}\nolimits} \left( {\bf{A}} \right)$ is the column-ordered vectorization of matrix ${\bf{A}}$, and ${{\mathop{\rm vec}\nolimits} ^{ - 1}}\left( {\bf{A}} \right)$ is the reverse operation of ${\mathop{\rm vec}\nolimits} \left( {\bf{A}} \right)$. ${{\cal O}}\left( \cdot \right)$ indicates the number of complex multiplications.

\section{System Model and Channel Model}
\subsection{System Model}
\begin{figure}[!t]
\centering
\includegraphics[width=15 cm]{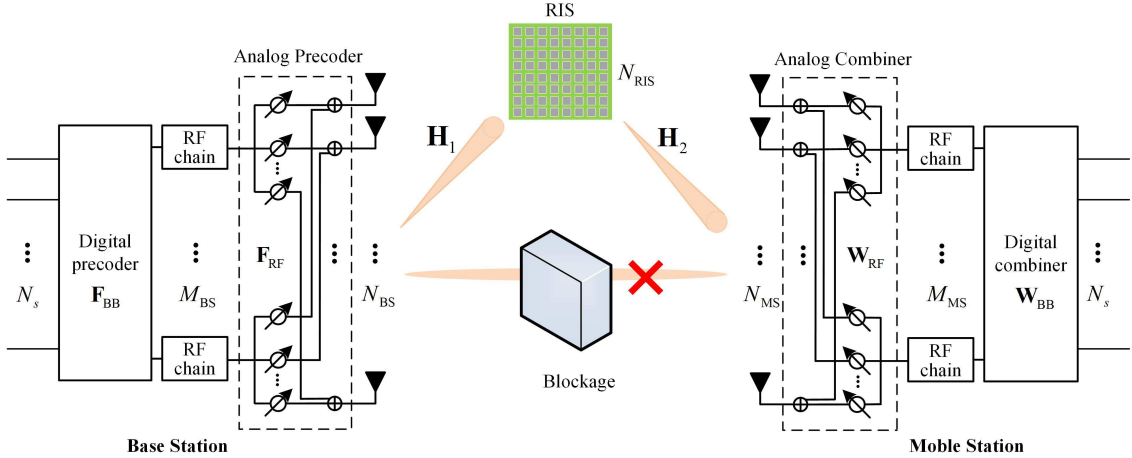}
\caption{Illustration of a RIS-empowered downlink THz MIMO system.}\label{system01}
\end{figure}


Considering a downlink THz MIMO system with hybrid beamforming architecture as shown in Fig. \,\ref{system01}, a BS employs ${{N_{{\mathop{\rm BS}\nolimits} }}}$ antennas to serve a MS equipped with ${{N_{{\mathop{\rm MS}\nolimits} }}}$ antennas. Since the LoS path between the BS and the MS is usually blocked by the obstacles, we suppose that the RIS is installed to assist this THz communication link, which consists of ${{N_{{\mathop{\rm RIS}\nolimits} }}}$ passive reflecting elements. In addition, a controller that connects the BS and the RIS is also required to realize the phase adjustment. Also, we assume that there are ${{M_{{\rm{BS}}}}}$ RF chains at BS side, and ${{M_{{\rm{MS}}}}}$ RF chains at MS side. Due to the serious power consumption of RF circuits, the number of antennas is larger than the number of the RF chains, i.e. ${{N_{{\rm{BS}}}} > {M_{{\rm{BS}}}}}$, ${{N_{{\rm{MS}}}} > {M_{{\rm{MS}}}}}$.
When the BS sends ${{N_s}}$ data streams ${{\bf{s}} \in {{\mathbb C}^{{N_s} \times 1}}}$ that satisfies ${\rm{E}}[{\bf{s}}{{\bf{s}}^H}] = \frac{1}{{{N_s}}}{{\bf{I}}_{{N_s}}}$, the MS employs ${{M_{{\rm{MS}}}}}$ RF chains to receive the processed signals. Thus, the received signal ${{{\bf{y}}_r} \in {{\mathbb C}^{{N_{{\mathop{s}\nolimits} }} \times 1}}}$ can be expressed as
\begin{align}\label{system_1}
{{\bf{y}}_r} = \sqrt \rho  {\bf{W}}_{{\mathop{\rm BB}\nolimits} }^H{\bf{W}}_{{\mathop{\rm RF}\nolimits} }^H{{\bf{H}}_2}{\bf{\Phi }}{{\bf{H}}_1}{{\bf{F}}_{{\mathop{\rm RF}\nolimits} }}{{\bf{F}}_{{\mathop{\rm BB}\nolimits} }}{\bf{s}} + {\bf{W}}_{{\mathop{\rm BB}\nolimits} }^H{\bf{W}}_{{\mathop{\rm RF}\nolimits} }^H{\bf{n}},
\end{align}
where ${\rho }$ is the transmission power; ${{{\bf{H}}_1} \in {{\mathbb C}^{{N_{{\mathop{\rm RIS}\nolimits} }} \times {N_{{\mathop{\rm BS}\nolimits} }}}}}$ indicates BS-RIS channel; ${{{\bf{H}}_2} \in {{\mathbb C}^{{N_{{\mathop{\rm MS}\nolimits} }} \times {N_{{\mathop{\rm RIS}\nolimits} }}}}}$ indicates RIS-MS channel; ${{{\bf{F}}_{{\mathop{\rm RF}\nolimits} }} \in {{\mathbb C}^{{N_{{\mathop{\rm BS}\nolimits} }} \times {M_{{\mathop{\rm BS}\nolimits} }}}}}$ (${{{\bf{F}}_{{\mathop{\rm BB}\nolimits} }} \in {{\mathbb C}^{{M_{{\mathop{\rm BS}\nolimits} }} \times {N_s}}}}$) denotes the analog (digital) precoding matrix; ${{{\bf{W}}_{{\mathop{\rm RF}\nolimits} }} \in {{\mathbb C}^{{N_{{\mathop{\rm MS}\nolimits} }} \times {M_{{\mathop{\rm MS}\nolimits} }}}}}$ (${{{\bf{W}}_{{\mathop{\rm BB}\nolimits} }} \in {{\mathbb C}^{{M_{{\mathop{\rm MS}\nolimits} }} \times {N_s}}}}$) denotes the analog (digital) combining matrix; ${{\bf{n}} \in {{\mathbb C}^{{N_{{\mathop{\rm MS}\nolimits} }} \times 1}}}$ represents the additive white Gaussian noise (AWGN) vector following the distribution of ${{\cal C}{\cal N}}\left( {{\bf{0}},{\delta ^2}{{\bf{I}}_{{N_{{\rm{MS}}}}}}} \right)$; and ${{\bf{\Phi }} = {\mathop{\rm diag}\nolimits} \left( {{{ { \mu {e^{j{\varphi _1}}}, \mu {e^{j{\varphi _2}}}, \cdots , \mu {e^{j{\varphi _{{N_{{\mathop{\rm RIS}\nolimits} }}}}}}} }}} \right)}$ is a diagonal matrix with the dimension of ${{N_{{\mathop{\rm RIS}\nolimits} }} \times {N_{{\mathop{\rm RIS}\nolimits} }}}$, respectively. Each entry $\left\{ { \mu {e^{j{\varphi _n}}}} \right\}_{n = 1}^{{N_{{\rm{RIS}}}}}$ of ${{\bf{\Phi }}}$ indicates the reflecting coefficient of a RIS element, and is composed of the reflecting amplitude ${\mu}$ and the phase shift $\left\{ {{\varphi _n}} \right\}_{n = 1}^{{N_{{\rm{RIS}}}}}$. Both ${\mu}$ and ${\varphi _n}$ are closely related with the hardware structure of RIS that will be introduced in Section III.


\subsection{Channel Model}
The RIS-enabled THz MIMO channel model contains ${{{\bf{H}}_1}}$, ${\bf{\Phi }}$ and ${{{\bf{H}}_2}}$, and the cascaded channel is denoted as ${{\bf{H}}_{\rm{e}}}{\rm{ = }}{{\bf{H}}_2}{\bf{\Phi }}{{\bf{H}}_1}$. We assume that both ${{{\bf{H}}_1}}$ and ${{{\bf{H}}_2}}$ consist of a LoS path and several non-line-of-sight (NLoS) paths, as we take the sparse nature of the THz channel into consideration. On the basic of geometric channel model \cite{system_01}, the BS-RIS channel ${{{\bf{H}}_1}}$ can be written as
\begin{align}\label{system_3}
\begin{array}{l}
{{\bf{H}}_1} = \sqrt {{N_{{\mathop{\rm BS}\nolimits} }}{N_{{\mathop{\rm RIS}\nolimits} }}} {\alpha _0}{G_t}{{\bf{a}}_{{\mathop{\rm RIS}\nolimits} }}\left( {\theta _{{\mathop{\rm RIS}\nolimits} ,0}^1,\theta _{{\mathop{\rm RIS}\nolimits} ,0}^2} \right) {\bf{a}}_{{\mathop{\rm BS}\nolimits} }^H\left( {\theta _{{\mathop{\rm BS}\nolimits} ,0}^1,\theta _{{\mathop{\rm BS}\nolimits} ,0}^2} \right) \\ \;\;\;\;\; + \sqrt {\frac{{{N_{{\mathop{\rm BS}\nolimits} }}{N_{{\mathop{\rm RIS}\nolimits} }}}}{L}} \sum\limits_{l = 1}^L {{\alpha _l}{G_t}} {{\bf{a}}_{{\mathop{\rm RIS}\nolimits} }}\left( {\theta _{{\mathop{\rm RIS}\nolimits} ,l}^1,\theta _{{\mathop{\rm RIS}\nolimits} ,l}^2} \right){\bf{a}}_{{\mathop{\rm BS}\nolimits} }^H\left( {\theta _{{\mathop{\rm BS}\nolimits} ,l}^1,\theta _{{\mathop{\rm BS}\nolimits} ,l}^2} \right),
\end{array}
\end{align}
where ${L}$ is the number of NLoS paths; ${{G_t}}$ is the transmitting antenna gain; ${\theta _{{\mathop{\rm RIS}\nolimits} }^1}$ $({\theta _{{\mathop{\rm RIS}\nolimits} }^2})$ denotes the angle of arrival (AoA) of ${{{\bf{H}}_1}}$ in the azimuth (elevation) domain; and ${\theta _{{\mathop{\rm BS}\nolimits} }^1}$ $({\theta _{{\mathop{\rm BS}\nolimits} }^2})$ denotes the angles of departure (AoD) of ${{{\bf{H}}_1}}$ in the azimuth (elevation) domain, respectively. Considering the large number of THz array antennas, the uniform planar array (UPA) structure is adopted as the array geometry. The normalized array response for the ${{N_x}{N_y}}$-element UPA on $xy$-plane can be expressed as
\begin{align}\label{system_4}
\begin{array}{l}
{{\bf{a}}_{{\rm{BS}}}}\left( {\theta _{{\rm{BS}}}^1,\theta _{{\rm{BS}}}^2} \right){\rm{ = }}\frac{{\rm{1}}}{{\sqrt {{N_{{\rm{BS}}}}} }}\left[ {1, \cdots ,{e^{\frac{{j2\pi d}}{\lambda }\left( {p\sin \left( {\theta _{{\rm{BS}}}^2} \right)\cos \left( {\theta _{{\rm{BS}}}^1} \right) + q\cos \left( {\theta _{{\rm{BS}}}^2} \right)} \right)}},} \right.\\
{\left. {\;\;\;\;\;\;\;\;\;\;\;\;\;\;\;\;\;\;\;\;\;\;\;\; \cdots ,{e^{\frac{{j2\pi d}}{\lambda }\left( {\left( {{N_x} - 1} \right)\sin \left( {\theta _{{\rm{BS}}}^2} \right)\cos \left( {\theta _{{\rm{BS}}}^1} \right) + \left( {{N_y} - 1} \right)\cos \left( {\theta _{{\rm{BS}}}^2} \right)} \right)}}} \right]^T},
\end{array}
\end{align}
where $p \in \left[ {0,{N_x - 1}} \right]$, $q \in \left[ {0,{N_y - 1}} \right]$ and ${{N_x}{N_y} = {N_{{\mathop{\rm BS}\nolimits} }}}$, respectively. The spacing of THz array antenna is $d = {\lambda  \mathord{\left/ {\vphantom {\lambda  2}} \right. \kern-\nulldelimiterspace} 2}$ where $\lambda$ is the incident wavelength. Similar to (\ref{system_4}), ${{{\bf{a}}_{{\mathop{\rm RIS}\nolimits} }}\left( {\theta _{{\mathop{\rm RIS}\nolimits} }^1,\theta _{{\mathop{\rm RIS}\nolimits} }^2} \right)}$ also employs the UPA structure, but the spacing of the RIS elements is the side length of each reflecting element. In addition, ${{\alpha _0}}$ is the LoS path gain of ${{{\bf{H}}_1}}$. As discussed in \cite{system_02}, ${{\alpha _0}}$ consists of spreading loss ${{\alpha _{{\mathop{\rm Spr}\nolimits} }}\left( f \right)}$ and molecular absorbing loss ${{\alpha _{{\mathop{\rm Abs}\nolimits} }}\left( f \right)}$, which can be expressed as
\begin{align}\label{system_5}
\begin{array}{l}
{\alpha _{\rm{0}}}{\rm{ = }}{\alpha _{{\rm{Spr}}}}\left( f \right) \cdot {\alpha _{{\rm{Abs}}}}\left( f \right) \cdot {e^{ - j2\pi f{\tau _{{\rm{Los}}}}}}= \frac{c}{{4\pi f{r_0}}} \cdot {e^{ - \frac{1}{2}\kappa \left( f \right){r_0}}} \cdot {e^{ - j2\pi f{\tau _{{\rm{Los}}}}}},
\end{array}
\end{align}
where ${c}$ is the speed of light; ${r_0}$ is the straight distance between the BS and the MS; ${{\tau _{{\mathop{\rm Los}\nolimits} }} = {{r_0} \mathord{\left/ {\vphantom {r c}} \right.
 \kern-\nulldelimiterspace} c}}$ is the time-of-arrival of the LoS path; and ${{\kappa \left( f \right)}}$ is the molecular absorbing coefficient. In addition, the channel gain ${{\alpha _l}}$ for the ${l}$th reflected path based on \cite{system_02} and \cite{system_02-1} can be written as
\begin{align}\label{system_6}
{\alpha _l} = \frac{c}{{4\pi  f  \left( {{r_1} + {r_2}} \right)}} \cdot {e^{ - \frac{1}{2}\kappa \left( f \right)\left( {{r_1} + {r_2}} \right)}} \cdot {e^{ - j2\pi f{\tau _{{\mathop{\rm Ref}\nolimits} }}}} \cdot \xi \left( f \right),
\end{align}
where $\xi \left( f \right)$ is the reflection coefficient of the reflecting materials (e.g., concrete, plastic, glass); ${{{r_1}}}$ is the distance between the transmitter and the reflecting material; ${{{r_2}}}$ is the distance between the receiver and the reflecting material; and ${{\tau _{{\mathop{\rm Ref}\nolimits} }} = {\tau _{{\mathop{\rm Los}\nolimits} }} + {{\left( {{r_1} + {r_2} - r} \right)} \mathord{\left/ {\vphantom {{\left( {{r_1} + {r_2} - r} \right)} c}} \right. \kern-\nulldelimiterspace} c}}$ is the time-of-arrival of the reflected path, respectively. Besides, the channel characteristics of ${{{\bf{H}}_2}}$ are identical to ${{{\bf{H}}_1}}$, so we can generate ${{{\bf{H}}_2}}$ in the same way. We assume that the channel estimation problem has been extensively studied in \cite{introduction_21,introduction_21-1,introduction_21-2,system03}, and these effective channel estimation methods can be well leveraged in this paper.

\section{Graphene Based Hardware Design of RIS}
In this section, an efficient graphene-based RIS is designed that realizes a wide range of amplitude response and phase response. In addition, the hardware design of the whole RIS is also accomplished, including the element arrangement, size selection in various situations and discrete phase distribution for a simplified hardware structure.

\subsection{Electric Properties of Graphene at THz Band}
In order to achieve the beam controllability, tunable components are embedded into the RIS elements \cite{yaojia_01}. Since the physical size of each reflecting element at THz band is too much tiny, the diodes and transistors can not be integrated into such a structure. In this case, graphene is a kind of appropriate material to facilitate the RIS with ultra-small size and tunable property.

Graphene is a two-dimensional material consisting of a single layer of carbon atoms. The conductivity of graphene can be altered through the applied voltage bias in a relatively wide range. Therefore, graphene provides various resonant states for each RIS element. According to \cite{yaojia_08}, the conductivity of graphene at THz band can be written as
\begin{align}\label{yaojia_01}
\sigma {\rm{ = }}\frac{{2{e^2}}}{{\pi {\hbar ^2}}}{k_B}T \cdot \ln \left[ {2\cosh \left( {\frac{{{E_F}}}{{2{k_B}T}}} \right)} \right]\frac{i}{{\omega  + i{\tau ^{ - 1}}}},
\end{align}
where ${e}$ is the elementary charge; ${\hbar }$ is the reduced Planck constant; ${{{k_B}}}$ is the Boltzmann constant; ${T}$ is the temperature; ${{{E_F}}}$ is the Fermi level; ${\tau }$ is the relaxation time; and ${\omega }$ is the angular frequency, respectively. It can be concluded that at a certain frequency point, the conductivity is only determined by Fermi level \cite{yaojia_09}. Then we can get the following expression as
\begin{align}\label{yaojia_02}
\left| {{E_F}} \right| = \hbar {\nu _F}\sqrt {\pi {n_{d}}} ,
\end{align}
where ${{\nu _F}}$ is the Fermi velocity and ${n_{d}}$ is the carrier density which can be expressed as
\begin{align}\label{yaojia_03}
n_{d} = \sqrt {n_0^2 + \alpha_c {{\left| {{V_{{\rm{CNP}}}} - {V_g}} \right|}^2}} ,
\end{align}
where ${{n_0}}$ is the residual carrier density and ${\alpha_c }$ is capacitivity related to the electrode. Besides, ${V_{{\rm{CNP}}}}$ is the compensating voltage, and ${{V_g}}$ is the applied voltage \cite{yaojia_10}. In summary, the conductivity of graphene can be continuously changed by the applied voltages, which is the foundation of forming an electrically controlled RIS.

\begin{figure}[!t]
\centering
\includegraphics[width=6cm]{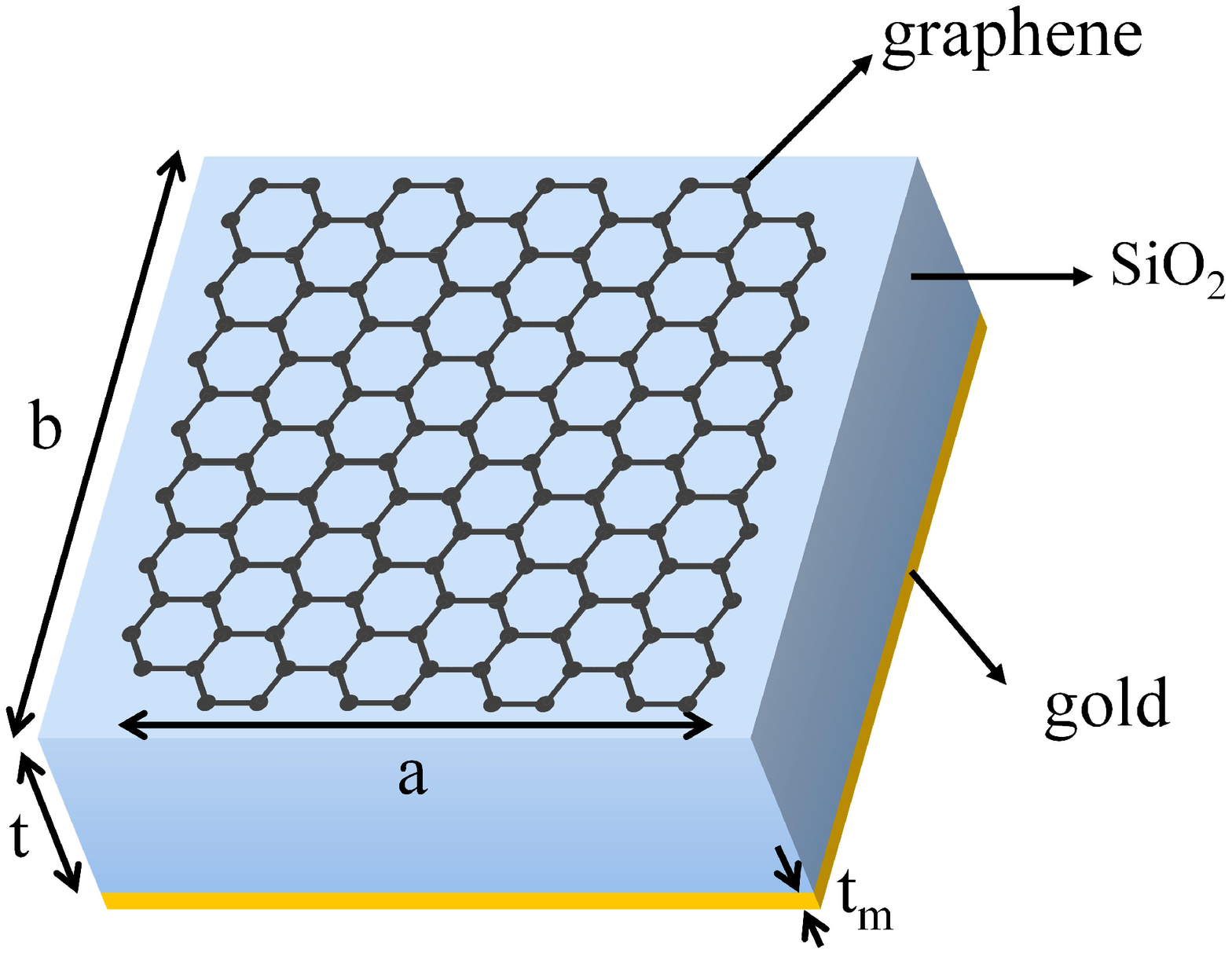}
\caption{Hardware structure of a RIS element that consists of graphene, quartz (substrate) and gold (ground plane) with ${a = 66{\mathop{\rm um}\nolimits} }$, ${b = 70{\mathop{\rm um}\nolimits} }$, ${t = 38{\mathop{\rm um}\nolimits} }$ and ${{t_m} = 1{\mathop{\rm um}\nolimits} }$.}\label{yaojia01}
\end{figure}

\subsection{Hardware Design of Graphene Based RIS}
The EM responses of the reflecting elements play an important role in the hardware structure of the RIS. Fig.\,\ref{yaojia01} shows a typical hardware design of a RIS element, which can be divided into three parts from top to the bottom: the graphene layer, the substrate and the metallic ground plane. The resonance model of this RIS architecture can be described as a Fabry-Perot cavity, where EM waves reflect back and forth between the top and the bottom surfaces. In addition, the resonance responses are caused by constructive or destructive interference of the multiple reflections \cite{yaojia_11}. In terms of such a reflecting element structure as shown in Fig.\,\ref{yaojia01}, the reflecting phase response based on \cite{yaojia_12} can be expressed as
\begin{align}\label{yaojia_04}
\varphi  = m\pi  - a{k_0}{\mathop{\rm Re}\nolimits} \left( {{n_{eff}}} \right),
\end{align}
where ${m}$ is an integer; ${a}$ is the width of graphene patch; ${{k_0}}$ is the wave number of free space; and ${{{n_{eff}}}}$ is the effective refraction index of the resonant structure, which is related to the effective permittivity ${{\varepsilon _{eff}}}$ of the graphene. In light of \cite{yaojia_13}, the parameter ${{\varepsilon _{eff}}}$ can be written as
\begin{align}\label{yaojia_05}
{\varepsilon _{eff}} = 1 + \frac{{i\sigma _c }}{{\omega {\varepsilon _0}{t_g}}},
\end{align}
where ${\sigma _c }$ is the conductivity and ${{{t_g}}}$ denotes the thickness of graphene. Combining (\ref{yaojia_04}) and (\ref{yaojia_05}), the phase response can be altered by the conductivity of graphene as well as the applied voltages.

\begin{figure*}[!t]
\centering
\includegraphics[width=16.2 cm]{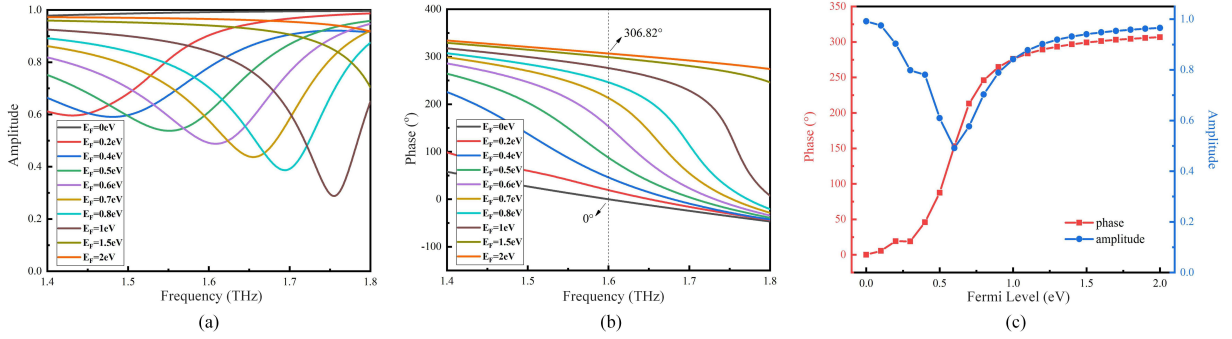}
\caption{Simulation results of the reflecting coefficient: (a) The response of reflecting amplitude; (b) The response of phase shift; (c) Phase response and reflecting amplitude versus Fermi level at 1.6 THz, where the phase response is ${\left[ {0^\circ ,306.82^\circ } \right]}$ with reflecting amplitude more than ${50\% }$.}\label{yaojia02}
\end{figure*}

The reflecting elements are simulated by leveraging the frequency domain solver in the simulation environment of CST Microwave Studio 2016. It is worth noting that a single reflecting element is unable to work since the miniature size causes the strong scattering. As a result, the boundary condition of the RIS is set as `unit cell' to mimic the repeated arrangement of the RIS elements. Fig.\,\ref{yaojia02} illustrates the reflecting coefficients from 1.4 THz to 1.8 THz with various Fermi levels. By combining Fig.\,\ref{yaojia02} (a) and Fig.\,\ref{yaojia02} (b), our designed RIS performs relatively stable broadband characteristics. However, a narrowband working mode of the RIS elements is selected in this paper where the center frequency is located at 1.6 THz.  Fig.\,\ref{yaojia02} (c) verifies that the amplitude efficiency of our designed reflecting element at 1.6 THz is more than ${50\% }$ and the phase response reaches to 306.82 degrees along with the chemical potential ranging from 0 ev to 2 eV. Moreover, the discrete phase shifts at 1.6 THz are also well-distributed with diverse Fermi levels, which lays the foundation for the bit quantization operation of phase shifts.

%

\begin{figure}[!t]
\centering
\includegraphics[width=11 cm]{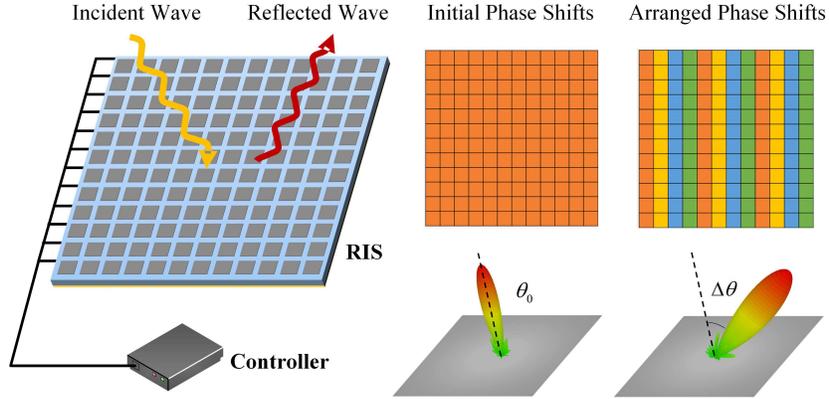}
\caption{Hardware design of the entire RIS device that consists of a RIS array and a controller. Various phase distributions represented by different color blocks are realized through applying different voltages to the RIS elements.}\label{yaojia03}
\end{figure}

Once the hardware structure of a single reflecting element is designed, the whole RIS is able to be accomplished by arranging massive reflecting elements closely in an array structure, as shown in Fig.\,\ref{yaojia03}. Then, various beam steering functions can be realized by controlling the phase response of all the RIS elements. But in practice, the appropriate number of the RIS elements needs to be carefully selected, which can make a better tradeoff between performance and deployment cost. According to Fig.\,\ref{yaojia02} (c), any expected phase shift within the phase response range can be obtained via applying the voltages continuously. However, the continuous phase control for each RIS element results in practical problems \cite{yaojia_14}, such as hardware complexity, power consumption, size limitation and the accuracy of phase control. To this end, the discrete phase shifts of the RIS elements are considered as the hardware structure in this paper.

For the combination of hardware design and system model, we firstly define the discrete phase set of each RIS element as ${{\cal F} = \left\{ {0,{{{\varphi _{\max }}} \mathord{\left/ {\vphantom {{{\varphi _{\max }}} {{2^b}}}} \right. \kern-\nulldelimiterspace} {{2^b}}}, \cdots ,{{({2^b} - 1){\varphi _{\max }}} \mathord{\left/ {\vphantom {{({2^b} - 1){\varphi _{\max }}} {{2^b}}}} \right. \kern-\nulldelimiterspace} {{2^b}}}} \right\}}$, where ${\varphi _{\max }}$ is the maximum phase response and ${b}$ is the bit quantization number. Then, we define the reflecting amplitude set as ${{\cal A}}{\rm{ = \{ }}{\mu _1}, \cdots ,{\mu _{\left| {{\cal F}} \right|}}{\rm{\} }}$, where $\left| {{\cal F}} \right| = {2^b}$. From Fig.\,\ref{yaojia02} (c) we can note that  when the distribution of the discrete phase shifts is determined, $\cal A$ can be acquired accordingly. In other words, there is a fixed mapping relationship between phase shift and reflecting amplitude for a specific hardware structure of RIS. Thus, the reflecting amplitude for each reflecting element can be further defined as $\bar \mu  = {{\left( {\sum\nolimits_{i = 1}^{|{{\cal F}}|} {{\mu _i}} } \right)} \mathord{\left/ {\vphantom {{\left( {\sum\nolimits_{i = 1}^{|{{\cal F}}|} {{\mu _i}} } \right)} {|{{\cal F}}|}}} \right. \kern-\nulldelimiterspace} {|{{\cal F}}|}}$, where the averaged amplitude $\bar \mu  \in \left[ {0.5,1} \right]$ is determined by parameter $b$ and the considered hardware structure of a RIS. Specifically, considering these practical constraints, the maximum phase shift and reflecting amplitude of RIS elements in this paper are set as 306.82 degrees and 0.8, respectively.


\section{Proposed Adaptive Gradient Descent Scheme}
With the given system model and hardware structure of RIS, the achievable rate of the RIS-assisted THz MIMO system can be written as
\begin{align}\label{system_6}
R = {\log _2}\left| {{{\bf{I}}_{{N_s}}} + \frac{\rho }{{{\delta ^2}{N_s}}}{{\left( {{{\bf{W}}^H}{\bf{W}}} \right)}^{ - 1}}{{\bf{W}}^H}} { \left( {{{\bf{H}}_2}{\bf{\Phi }}{{\bf{H}}_1}} \right){\bf{F}}{{\bf{F}}^H}{{\left( {{{\bf{H}}_2}{\bf{\Phi }}{{\bf{H}}_1}} \right)}^H}{\bf{W}}} \right|,
\end{align}
where ${{\bf{F}} = {{\bf{F}}_{{\mathop{\rm RF}\nolimits} }}{{\bf{F}}_{{\mathop{\rm BB}\nolimits} }}}$ and ${{\bf{W}} = {{\bf{W}}_{{\mathop{\rm RF}\nolimits} }}{{\bf{W}}_{{\mathop{\rm BB}\nolimits} }}}$.
To convert the optimization problem (\ref{system_6}) into a more evident form, the achievable rate maximization problem for obtaining the optimal hybrid beamforming and phase shifts can be rewritten as
\begin{align}\label{system_7}
\begin{array}{l}
\left( {{{\bf{\Phi }}^{{\rm{opt }}}},{{\bf{W}}^{{\rm{opt }}}},{{\bf{F}}^{{\rm{opt }}}}} \right) = \mathop {\arg \max }\limits_{{\bf{\Phi }},{\bf{W}},{\bf{F}}} R\\
s.t.\;\;{\rm{ }}{\varphi _n} \in {{\cal F}},\;{\varphi _{\max }} = {306.82^{\rm{o}}},\;\forall n = 1, \ldots ,{N_{{\rm{RIS}}}},\\
\;\;\;\;\;\;\;\left| {\bar \mu {e^{j{\varphi _n}}}} \right| = 0.8,\;\forall n = 1, \ldots ,{N_{{\rm{RIS}}}},\\
\;\;\;\;\;\;\;\left\| {\bf{F}} \right\|_F^2 = {N_s},
\end{array}
\end{align}
where the first constraint is from our designed discrete phase response and the second constraint stems from our designed reflecting amplitude. Due to the non-convex and discrete constraints, it is challenging to directly solve the problem (\ref{system_7}). Fortunately, one available way of settling such an optimization problem is to first design hybrid beamforming ${{\bf{F}}}$ and ${{\bf{W}}}$, and then optimize the phase shift matrix ${{\bf{\Phi }}}$, respectively.

Given a fixed ${{\bf{\Phi }}}$, the singular value decomposition (SVD) of the cascaded channel ${{\bf{H}}_{\rm{e}}}$ can be written as
\begin{align}\label{wenjie_3}
\begin{array}{l}
{{\bf{H}}_{\mathop{\rm e}\nolimits} } = {\bf{U\Lambda }}{{\bf{V}}^H} = \left[ {{{\bf{U}}_1},{{\bf{U}}_2}} \right]\left[ {\begin{array}{*{20}{c}}
{{{\bf{\Lambda }}_1}}&{\bf{0}}\\
{\bf{0}}&{{{\bf{\Lambda }}_2}}
\end{array}} \right]{\left[ {{{\bf{V}}_1},{{\bf{V}}_2}} \right]^H},
\end{array}
\end{align}
where ${{\bf{U}}}$ is a ${{N_{{\mathop{\rm MS}\nolimits} }} \times G}$ unitary matrix; ${{\bf{\Lambda }}}$ is a ${G \times G}$ dimensional matrix; ${{\bf{V}}}$ is a ${{N_{{\mathop{\rm BS}\nolimits} }} \times G}$ unitary matrix; ${{\bf{U}}_1}$ is a ${N_{{\rm{MS}}}} \times N_s$ submatrix of ${{\bf{U}} = \left[ {{{\bf{U}}_1},{{\bf{U}}_2}} \right]}$; and ${{\bf{V}}_1}$ is a ${N_{{\rm{BS}}}} \times N_s$ submatrix of ${{\bf{V}} = \left[ {{{\bf{V}}_1},{{\bf{V}}_2}} \right]}$, in which $G \buildrel \Delta \over = rank\left( {{{\bf{H}}_{\rm{e}}}} \right)$. Especially, ${{\bf{\Lambda }}_1}$ and ${{\bf{\Lambda }}_2}$ are $N_s \times N_s$ and $(G-N_s) \times (G-N_s)$ diagonal matrices with the singular values arranged in a decreasing order.

According to \cite{introduction_33}, the optimal combiner and precoder are ${{\bf{W}^{{\rm{opt}}}} = {{\bf{U}}_1}}$ and ${{\bf{F}^{{\rm{opt}}}} = {{\bf{V}}_1} }$ which satisfies ${\left\| {\bf{F}} \right\|_F^2 = {N_s}}$. Thus, the achievable rate ${R}$ in (\ref{system_6}) can be rewritten as
\begin{align}\label{wenjie_4}
R = {\log _2}\left| {{{\bf{I}}_{{N_s}}} + \frac{\rho }{{{\delta ^2}{N_s}}}{{\bf{\Lambda }}_1}{\bf{\Lambda }}_1^H} \right|.
\end{align}


Furthermore, (\ref{wenjie_4}) can be further simplified as
\begin{align}\label{wenjie_5}
\begin{array}{l}
R = {\log _2}\left| {{{\bf{I}}_{{N_s}}} + \frac{\rho }{{{\delta ^2}{N_s}}}{\bf{\Lambda }}_1^2} \right|\\
\;\;\; \mathop  \le \limits^{(a)} {N_s}{\log _2}\left( {1 + \frac{\rho }{{{\delta ^2}{N_s}}}{\rm{tr}}\left( {{\bf{\Lambda }}_1^2} \right)} \right)\\
\;\;\; \mathop  \le \limits^{(b)} {N_s}{\log _2}\left( {1 + \frac{\rho }{{{\delta ^2}{N_s}}}{\rm{tr}}\left( {{{\bf{H}}_{\rm{e}}}{\bf{H}}_{\rm{e}}^H} \right)} \right),
\end{array}
\end{align}
where $(a)$ comes from Jensen's inequality and $(b)$ takes the mark of equality when $N_s = G$. Therefore, the optimization problem (\ref{system_7}) can be formulated as
\begin{align}\label{wenjie_5-1}
\begin{array}{l}
{{\bf{\Phi }}^{{\rm{opt }}}} = \mathop {\arg \max }\limits_{\bf{\Phi }} \;\; {\rm{tr}}\left( {{{\bf{H}}_{\rm{e}}}{\bf{H}}_{\rm{e}}^H} \right)\\
s.t.\;\;{\rm{ }}{\varphi _n} \in {{\cal F}},\;{\varphi _{\max }} = {306.82^{\rm{o}}},\;\forall n = 1, \ldots ,{N_{{\rm{RIS}}}},\\
\;\;\;\;\;\;\;\left| {\bar \mu {e^{j{\varphi _n}}}} \right| = 0.8,\;\forall n = 1, \ldots ,{N_{{\rm{RIS}}}}.
\end{array}
\end{align}

Nevertheless, the objective (\ref{wenjie_5-1}) is still a constrained optimization problem, since ${\bf{\Phi }}$ possesses discrete phase shifts and constant-magnitude entries. Let us define ${{\bm{\varphi }} \buildrel \Delta \over = \left[ {{{{{\varphi _1},{\varphi _2},...,{\varphi _{{N_{{\mathop{\rm RIS}\nolimits} }}}}} }}} \right]}$, and then consider ${{\bf{\Phi }}}$ as a function of ${{\bm{\varphi }}}$ where ${{\bf{\Phi }}\left( \bm{\varphi }  \right) = } {\left[ {{e^{j{\varphi _1}}},{e^{j{\varphi _2}}}, \cdots ,{e^{j{\varphi _{{N_{{\mathop{\rm RIS}\nolimits} }}}}}}} \right]}$. Given this, (\ref{wenjie_5-1}) can be reformulated as
\begin{align}\label{wenjie_11}
\mathop {\min }\limits_{\bm{\varphi }}  - {\mathop{\rm Tr}\nolimits} \left( {{{\bf{H}}_2}{\bf{\Phi }}\left( {\bm{\varphi }} \right){{\bf{H}}_1}{{\left( {{{\bf{H}}_2}{\bf{\Phi }}\left( {\bm{\varphi }} \right){{\bf{H}}_1}} \right)}^H}} \right) = \mathop {\min }\limits_{\bm{\varphi }} f({\bm{\varphi }}),
\end{align}
where ${{\mathop{f}\nolimits} \left(  \cdot  \right)}$ is a compound function relating to ${{\bm{\varphi }}}$, which can be further written as
\begin{align}\label{wenjie_12}
\begin{array}{l}
f({\bm{\varphi }}) \mathop  = \limits^{(c)}  - \left\| {{{\bf{H}}_2}{\bf{\Phi }}({\bm{\varphi }}){{\bf{H}}_1}} \right\|_F^2\\
\;\;\;\;\;\;\;\; \mathop= \limits^{(d)}  - \left\| {{\mathop{\rm vec}\nolimits} \left( {{{\bf{H}}_2}{\bf{\Phi }}({\bm{\varphi }}){{\bf{H}}_1}} \right)} \right\|_F^2\\
\;\;\;\;\;\;\;\; =  - \left\| {\left( {{\bf{H}}_1^T \otimes {{\bf{H}}_2}} \right){\mathop{\rm vec}\nolimits} \left( {{\bf{\Phi }}({\bm{\varphi }})} \right)} \right\|_F^2\\
\;\;\;\;\;\;\;\; =  - {\mathop{\rm vec}\nolimits} {\left( {{\bf{\Phi }}({\bm{\varphi }})} \right)^H}{\left( {{\bf{H}}_1^T \otimes {{\bf{H}}_2}} \right)^H}\left( {{\bf{H}}_1^T \otimes {{\bf{H}}_2}} \right){\mathop{\rm vec}\nolimits} \left( {{\bf{\Phi }}({\bm{\varphi }})} \right),
\end{array}
\end{align}
where equality $(c)$ stems from the Frobenius norm operation and equality $(d)$ is the vectorization operator. In addition, ${{\left( {{\bf{H}}_1^T \otimes {{\bf{H}}_2}} \right)^H}\left( {{\bf{H}}_1^T \otimes {{\bf{H}}_2}} \right)}$ is independent of the variable matrix ${{\bm{\varphi }}}$. In order to further simplify (\ref{wenjie_12}), we define that
\begin{align}\label{wenjie_13}
{\bf{A}} \buildrel \Delta \over =  - {\left( {{\bf{H}}_1^T \otimes {{\bf{H}}_2}} \right)^H}\left( {{\bf{H}}_1^T \otimes {{\bf{H}}_2}} \right),
\end{align}
\begin{align}\label{wenjie_14}
{\bf{x}} \buildrel \Delta \over = {\mathop{\rm vec}\nolimits} \left( {{\bf{\Phi }}\left( {\bm{\varphi }} \right)} \right),
\end{align}
where ${{\bf{A}} \in {{\mathbb C}^{{N_{{\mathop{\rm BS}\nolimits} }}{N_{{\mathop{\rm MS}\nolimits} }} \times N_{{\mathop{\rm RIS}\nolimits} }^2}}}$ is a conjugate symmetric positive definite matrix, and ${\bf{x}} \in {^{N_{{\rm{RIS}}}^2 \times 1}}$ is a sparse vector. Then, (\ref{wenjie_12}) can be written as
\begin{align}\label{wenjie_15}
\mathop {\min }\limits_{\bm{\varphi }} {\mathop{f}\nolimits} \left( {\bm{\varphi }} \right) = \mathop {\min }\limits_{\bm{\varphi }} {{\bf{x}}^H}{\bf{Ax}}.
\end{align}


Furthermore, by substituting (\ref{wenjie_13}) and (\ref{wenjie_14}) into (\ref{wenjie_15}), ${{{\bf{x}}^H}{\bf{Ax}}}$ can be written as
\begin{align}\label{wenjie_16}
\begin{array}{l}
{{\bf{x}}^H}{\bf{Ax}} = \left[ {\bar \mu {e^{ - j{\varphi _1}}},{{\bf{0}}_{1 \times {N_{{\rm{RIS}}}}}},\bar \mu {e^{ - j{\varphi _2}}},{{\bf{0}}_{1 \times {N_{{\rm{RIS}}}}}}, \cdots ,\bar \mu {e^{ - j{\varphi _{{N_{{\rm{RIS}}}}}}}}} \right]\\
\;\;\;\;\;\;\;\;\;\;\; \times {\bf{A}}{\left[ {\bar \mu {e^{j{\varphi _1}}},{{\bf{0}}_{1 \times {N_{{\rm{RIS}}}}}},\bar \mu {e^{j{\varphi _2}}},{{\bf{0}}_{1 \times {N_{{\rm{RIS}}}}}}, \cdots ,\bar \mu {e^{j{\varphi _{{N_{{\rm{RIS}}}}}}}}} \right]^T}\\
\;\;\;\;\;\;\;\;\;\;\; = {{\bar \mu }^2}\sum\limits_{p = 1}^{{N_{{\rm{RIS}}}}} {\sum\limits_{q = 1}^{{N_{{\rm{RIS}}}}} {{e^{j\left( {{\varphi _q} - {\varphi _p}} \right)}}} } {{\bf{A}}_{(p - 1){N_{{\rm{RIS}}}} + p,(q - 1){N_{{\rm{RIS}}}} + q}},
\end{array}
\end{align}
where ${{{\bf{A}}_{\left( {p - 1} \right){N_{{\mathop{\rm RIS}\nolimits} }} + p,\left( {q - 1} \right){N_{{\mathop{\rm RIS}\nolimits} }} + q}}}$ represents ${\left( {\left( {p - 1} \right){N_{{\mathop{\rm RIS}\nolimits} }} + p,\left( {q - 1} \right){N_{{\mathop{\rm RIS}\nolimits} }} + q} \right)}$th entry of ${{\bf{A}}}$ and ${{{\bf{0}}_{1 \times {N_{{\rm{RIS}}}}}}}$ is a ${1 \times  {{N_{{\mathop{\rm RIS}\nolimits} }}} }$ all-zero vector. Since ${{{\bf{A}}^H} = {\bf{A}}}$, each diagonal entry of ${{\bf{A}}}$ is a real value and ${{{\bf{A}}_{i,j}} + {{\bf{A}}_{j,i}} = 2{\cal R}\left\{ {{{\bf{A}}_{i,j}}} \right\}}$, ${\forall i,j = 1, \cdots ,{N_{{\mathop{\rm RIS}\nolimits} }},i \ne j}$. Therefore, (\ref{wenjie_16}) can be further simplified as
\begin{align}\label{wenjie_17}
\begin{array}{l}
{{\bf{x}}^H}{\bf{Ax}} = {{\bar \mu }^2}\sum\limits_{p = 1}^{{N_{{\rm{RIS}}}}} {{{\bf{A}}_{(p - 1){N_{{\rm{RIS}}}} + p,(p - 1){N_{{\rm{RIS}}}} + p}}} + 2{{\bar \mu }^2}{{\cal R}}\left\{ {\sum\limits_{p = 1}^{{N_{{\rm{RIS}}}}} {\sum\limits_{q > p}^{{N_{{\rm{RIS}}}}} {{e^{j\left( {{\varphi _q} - {\varphi _p}} \right)}}} } {{\bf{A}}_{(p - 1){N_{{\rm{RIS}}}} + p,(q - 1){N_{{\rm{RIS}}}} + q}}} \right\}.
\end{array}
\end{align}

We consider $\left\{ {{\varphi _n}} \right\}_{n = 1}^{{N_{{\rm{RIS}}}}}$ as continuous phase shifts and thus ${{\bm{\varphi }}}$ is unconstrained. Given this, the ${n}$th element of the gradient vector ${{\nabla _{\bm{\varphi }}}{\mathop{f}\nolimits} \left( {\bm{\varphi }} \right)}$ can be expressed as
\begin{align}\label{wenjie_18}
\begin{array}{l}
\frac{{\partial f({\bm{\varphi }})}}{{\partial {\varphi _n}}} = \frac{{\partial \left( {{{\bf{x}}^H}{\bf{Ax}}} \right)}}{{\partial {\varphi _n}}}\\
 = 2{{\bar \mu }^2}{{\cal R}}\left\{ { - j{e^{ - j{\varphi _n}}}\sum\limits_{q > n}^{{N_{{\rm{RIS}}}}} {{e^{j{\varphi _q}}}} {{\bf{A}}_{(n - 1){N_{{\rm{RIS}}}} + n,(q - 1){N_{{\rm{RIS}}}} + q}}} { + j{e^{j{\varphi _n}}}\sum\limits_{n > p}^{{N_{{\rm{RIS}}}}} {{e^{ - j{\varphi _p}}}} {{\bf{A}}_{(p - 1){N_{{\rm{RIS}}}} + p,(n - 1){N_{{\rm{RIS}}}} + n}}} \right\}.
\end{array}
\end{align}

After computing all the $\partial f({\bm{\varphi }})/\partial {\varphi _n},\forall n = 1,2, \cdots ,{N_{{\rm{RIS}}}}$, ${{\nabla _{\bm{\varphi }}}{\mathop{f}\nolimits} \left( {\bm{\varphi }} \right)}$ can be expressed as
\begin{align}\label{wenjie_19}
{\nabla _{\bm{\varphi }}}{\mathop{f}\nolimits} \left( {\bm{\varphi }} \right) = {\left[ {\frac{{\partial {\mathop{f}\nolimits} \left( {\bm{\varphi }} \right)}}{{\partial {\varphi _1}}},\frac{{\partial {\mathop{f}\nolimits} \left( {\bm{\varphi }} \right)}}{{\partial {\varphi _2}}}, \cdots ,\frac{{\partial {\mathop{f}\nolimits} \left( {\bm{\varphi }} \right)}}{{\partial {\varphi _{{N_{{\mathop{\rm RIS}\nolimits} }}}}}}} \right]^T}.
\end{align}

On basis of the gradient direction ${{\nabla _{\bm{\varphi }}}{\mathop{f}\nolimits} \left( {\bm{\varphi }} \right)}$, the objective function ${{\mathop{f}\nolimits} \left( {\bm{\varphi }} \right)}$ is able to descend by replacing ${{\bm{\varphi }}}$ with ${{\bm{\varphi }} - \lambda {\mathop{\rm diag}\nolimits} \left( {{\nabla _{\bm{\varphi }}}{\mathop{f}\nolimits} \left( {\bm{\varphi }} \right)} \right)}$, where ${\lambda }$ is the iterative step size. During the ${i}$th iteration, the updated ${{{\bm{\varphi }}^{\left( {i + 1} \right)}}}$ and the renewed ${{\mathop{f}\nolimits} \left( {{{\bm{\varphi }}^{i + 1}}} \right)}$ can be respectively written as
\begin{align}\label{wenjie_20}
{{\bm{\varphi }}^{i + 1}} = {{\bm{\varphi }}^i} - \lambda {\nabla _{\bm{\varphi }}}f\left( {{{\bm{\varphi }}^i}} \right),
\end{align}
\begin{align}\label{wenjie_21}
f\left( {{{\bm{\varphi }}^{i + 1}}} \right) = f\left\{ {{{\bm{\varphi }}^i} - \lambda {\nabla _{\bm{\varphi }}}f\left( {{{\bm{\varphi }}^i}} \right)} \right\}.
\end{align}

Since the C-GD algorithm obtains the fixed step size ${{\lambda}}$ by simulation experiment \cite{yanlongfei_01}, it suffers from high complexity and low efficiency. To this end, a novel A-GD algorithm is developed to determine the dynamic step size ${{\lambda ^i}}$. Specifically, we firstly expand (\ref{wenjie_21}) as
\begin{align}\label{wenjie_22}
\begin{array}{l}
{\mathop{f}\nolimits} \left( {{{\bm{\varphi }}^{i + 1}}} \right) = {\left( {{{\bf{x}}^{\left( {i + 1} \right)}}} \right)^H}{\bf{A}}{{\bf{x}}^{\left( {i + 1} \right)}}\\
\;\;\;\;\;\;\;\;\;\;\;\;\; = {{\bar \mu }^2}\sum\limits_{q = 1}^{{N_{{\rm{RIS}}}}} {{e^{j\left( {\varphi _q^i + \Delta \varphi _q^i} \right)}}} \sum\limits_{p = 1}^{{N_{{\rm{RIS}}}}} {\left( {{e^{ - j\left( {\varphi _p^i + \Delta \varphi _p^i} \right)}}{{\bf{A}}_{(p - 1){N_{{\rm{RIS}}}} + p,(q - 1){N_{{\rm{RIS}}}} + q}}} \right)} \\
\;\;\;\;\;\;\;\;\;\;\;\;\; = {{\bar \mu }^2}\sum\limits_{p = 1}^{{N_{{\rm{RIS}}}}} {\sum\limits_{q = 1}^{{N_{{\rm{RIS}}}}} {\left( {{e^{j\left( {\varphi _q^i + \Delta \varphi _q^i} \right) - j\left( {\varphi _p^i + \Delta \varphi _p^i} \right)}}{{\bf{A}}_{(p - 1){N_{{\rm{RIS}}}} + p,(q - 1){N_{{\rm{RIS}}}} + q}}} \right)} },
\end{array}
\end{align}
where ${\Delta \varphi _n^i =  - {\lambda ^i}{{\partial {\mathop{f}\nolimits} \left( {{{\bm{\varphi }}^i}} \right)} \mathord{\left/
 {\vphantom {{\partial {\mathop{f}\nolimits} \left( {{{\bf{\varphi }}^i}} \right)} {\partial \varphi _n^i}}} \right.
 \kern-\nulldelimiterspace} {\partial \varphi _n^i}}}$ for ${n = 1,2, \cdots ,{N_{{\mathop{\rm RIS}\nolimits} }}}$. By using ${{\bf{A}} = {{\bf{A}}^H}}$, ${{\mathop{f}\nolimits} \left( {{{\bm{\varphi }}^{i + 1}}} \right)}$ can be further simplified as
\begin{align}\label{wenjie_23}
\begin{array}{l}
{\mathop{f}\nolimits} \left( {{{\bm{\varphi }}^{i + 1}}} \right) = {{\bar \mu }^2}\sum\limits_{p = 1}^{{N_{{\mathop{\rm RIS}\nolimits} }}} {{{\bf{A}}_{\left( {p - 1} \right){N_{{\mathop{\rm RIS}\nolimits} }} + p,\left( {p - 1} \right){N_{{\mathop{\rm RIS}\nolimits} }} + p}}} \\
\;\;\;\;\;\;\;\;\;\;\;\;\; + 2{{\bar \mu }^2}{{\cal R}}\left\{ {\sum\limits_{p = 1}^{{N_{{\rm{RIS}}}}} {\sum\limits_{q > p}^{{N_{{\rm{RIS}}}}} {\left\{ {{e^{j\left( {\varphi _q^i - \varphi _p^i} \right)}}{e^{j{\lambda ^i}\left( {\frac{{\partial f\left( {{{\bm{\varphi }}^i}} \right)}}{{\partial \varphi _p^i}} - \frac{{\partial f\left( {{{\bm{\varphi }}^i}} \right)}}{{\partial \varphi _q^i}}} \right)}}{{\bf{A}}_{(p - 1){N_{{\rm{RIS}}}} + p,(q - 1){N_{{\rm{RIS}}}} + q}}} \right\}} } } \right\}.
\end{array}
\end{align}

According to (\ref{wenjie_23}), ${{\mathop{f}\nolimits} \left( {{{\bm{\varphi }}^{i + 1}}} \right)}$ is only determined by ${{\lambda ^i}}$ during the $(i+1)$th iteration process. To minimize ${{\mathop{f}\nolimits} \left( {{{\bm{\varphi }}^{i + 1}}} \right)}$, we need to search a much better ${{\lambda ^i}}$ and the corresponding optimization problem can be expressed as
\begin{align}\label{wenjie_24}
\begin{array}{l}
{\lambda ^i} = \arg \mathop {\min }\limits_{{\lambda ^i}} f\left( {{{\bm{\varphi }}^{i + 1}}} \right)\\
\;\;\;\; = \arg \mathop {\min }\limits_{{\lambda ^i}} {{\bar \mu }^2} {\cal R}\left\{ {\mathop \sum \limits_{p = 1}^{{N_{{\rm{RIS}}}}} \mathop \sum \limits_{q > p}^{{N_{{\rm{RIS}}}}} {e^{j\left( {\varphi _q^i - \varphi _p^i} \right)}}{e^{j{\lambda ^i}\left( {\frac{{\partial f\left( {{{\bm{\varphi }}^i}} \right)}}{{\partial \varphi _p^i}} - \frac{{\partial f\left( {{{\bm{\varphi }}^i}} \right)}}{{\partial \varphi _q^i}}} \right)}} } {{{\bf{A}}_{(p - 1){N_{{\rm{RIS}}}} + p,(q - 1){N_{{\rm{RIS}}}} + q}}} \right\}.
\end{array}
\end{align}

\begin{algorithm}[!t]
	\caption{Proposed A-GD Algorithm}
	\begin{algorithmic}[1]
		\REQUIRE ${{{\bf{H}}_{\rm{1}}}}$, ${{{\bf{H}}_{\rm{2}}}}$, ${\rho }$, ${{N_s}}$, ${{\delta ^2}}$, ${{\cal F}}$, $b$, ${\varphi _{\max }}$, ${I}$,
		\STATE Initialize ${{\bf{A}} =  - {\left( {{\bf{H}}_1^T \otimes {{\bf{H}}_2}} \right)^H}\left( {{\bf{H}}_1^T \otimes {{\bf{H}}_2}} \right)}$, ${{{\mathop{f}\nolimits} _{\max }} = 0}$, ${{\bf{W}^{{\rm{opt}}}} = {{\bf{U}}_1}}$,  ${{\bf{F}^{{\rm{opt}}}} = {{\bf{V}}_1} }$,\\ ${{{\bm{\varphi }}^0} = {{\bf{0}}_{{N_{{\mathop{\rm RIS}\nolimits} }} \times 1}}}$, ${{\bf{\Phi }}} = \bar \mu {{\bf{I}}_{{N_{{\rm{RIS}}}} \times {N_{{\rm{RIS}}}}}}$, ${{{\bf{x}}^0} = {\mathop{\rm vec}\nolimits} \left( {{\bf{\Phi }}\left( {{{\bm{\varphi }}^0}} \right)} \right)}$, ${i = 0}$,
        \STATE \textbf{while} ${i \le I}$ \textbf{do}
        \STATE ${{}}$ ${{}}$ Calculate the gradient vector ${{\nabla _{\bm{\varphi }}}{\mathop{f}\nolimits} \left( {{{\bm{\varphi }}^i}} \right)}$ in (\ref{wenjie_19}),
        \STATE ${{}}$ ${{}}$ Compute the step size ${{\lambda ^i}}$ in (\ref{wenjie_29}),
        \STATE ${{}}$ ${{}}$ Update ${{{\bm{\varphi }}^{i + 1}} = {{\bm{\varphi }}^i} - {\lambda ^i} {{\nabla _{\bm{\varphi }}}{\mathop{f}\nolimits} \left( {{{\bm{\varphi }}^i}} \right)} }$, ${{{\bf{x}}^{i + 1}} = {\mathop{\rm vec}\nolimits} \left( {{\bf{\Phi }}\left( {{{\bm{\varphi }}^i}} \right)} \right)}$,
        \STATE ${{}}$ ${{}}$ Renew ${{\mathop{f}\nolimits} \left( {{{\bm{\varphi }}^{i + 1}}} \right) = {\mathop{f}\nolimits} \left( {{{\bm{\varphi }}^i} - {\lambda ^i} {{\nabla _{\bm{\varphi }}}{\mathop{f}\nolimits} \left( {{{\bm{\varphi }}^i}} \right)}} \right)}$,
        \STATE ${{}}$ ${{}}$ \textbf{if} ${{\mathop{f}\nolimits} \left( {{{\bm{\varphi }}^{i + 1}}} \right) > {{\mathop{f}\nolimits} _{\max }}}$ \textbf{do}
        \STATE ${{}}$ ${{}}$ ${{}}$ ${{}}$ ${{{\mathop{f}\nolimits} _{\max }} = {\mathop{f}\nolimits} \left( {{{\bm{\varphi }}^{i + 1}}} \right)}$, ${{{\bm{\varphi }}^{{\mathop{\rm opt}\nolimits} }} = {{\bm{\varphi }}^{i + 1}}}$,
        \STATE ${{}}$ ${{}}$ \textbf{end if}
        \STATE ${{}}$ ${{}}$ ${i = i + 1}$,
        \STATE \textbf{end while}
        \STATE Map each entry of ${{{\bm{\varphi }}^{{\mathop{\rm opt}\nolimits} }}}$ into discrete phase shifts from ${{\cal F}}$,
        \STATE Calculate ${{{\bm{\Phi }}^{{\mathop{\rm opt}\nolimits} }} = {\mathop{\rm diag}\nolimits} \left( {\exp \left( {{{\bm{\varphi }}^{{\mathop{\rm opt}\nolimits} }}} \right)} \right)}$ and ${R}$ in (\ref{system_6}).
        \ENSURE ${{{\bf{\Phi }}^{{\mathop{\rm opt}\nolimits} }}}$, ${R}$
	\end{algorithmic}
\end{algorithm}

In order to optimize (\ref{wenjie_24}), the A-GD algorithm replaces ${{e^{j{\lambda ^i}\left[ {{{\partial {\mathop{f}\nolimits} \left( {{{\bm{\varphi }}^i}} \right)} \mathord{\left/ {\vphantom {{\partial {\mathop{f}\nolimits} \left( {{{\bm{\varphi }}^i}} \right)} {\partial \varphi _p^i}}} \right. \kern-\nulldelimiterspace} {\partial \varphi _p^i}} - {{\partial {\mathop{f}\nolimits} \left( {{{\bm{\varphi }}^i}} \right)} \mathord{\left/ {\vphantom {{\partial {\mathop{f}\nolimits} \left( {{{\bf{\varphi }}^i}} \right)} {\partial \varphi _q^i}}} \right.
 \kern-\nulldelimiterspace} {\partial \varphi _q^i}}} \right]}}}$ with the second-order Taylor expansion formulation. Thus, (\ref{wenjie_24}) can be approximated as
\begin{align}\label{wenjie_25}
\begin{array}{l}
{\lambda ^i} \approx \arg \mathop {\min }\limits_{{\lambda ^i}} {{\bar \mu }^2} {\cal R}\left\{ {\mathop \sum \limits_{p = 1}^{{N_{{\rm{RIS}}}}} \mathop \sum \limits_{q > p}^{{N_{{\rm{RIS}}}}} {e^{j\left( {\varphi _q^i - \varphi _p^i} \right)}}{{\bf{A}}_{(p - 1){N_{{\rm{RIS}}}} + p,(q - 1){N_{{\rm{RIS}}}} + q}} } \right.\\
\;\;\;\;\;\;\;\;\;\;\;\;\;\;\;\;\;\;\;\;\;\;\;\; \left. {\times  \left[ {1 + j{\lambda ^i}\left( {\frac{{\partial f\left( {{{\bm{\varphi }}^i}} \right)}}{{\partial \varphi _p^i}} - \frac{{\partial f\left( {{{\bm{\varphi }}^i}} \right)}}{{\partial \varphi _q^i}}} \right) + \frac{{{{\left( {j{\lambda ^i}} \right)}^2}}}{2}{{\left( {\frac{{\partial f\left( {{{\bm{\varphi }}^i}} \right)}}{{\partial \varphi _p^i}} - \frac{{\partial f\left( {{{\bm{\varphi }}^i}} \right)}}{{\partial \varphi _q^i}}} \right)}^2}} \right]} \right\}\\
\;\;\;\; = \arg \mathop {\min }\limits_{{\lambda ^i}} {C_0} + {C_1}{\lambda ^i} + {C_2}{\left( {{\lambda ^i}} \right)^2},
\end{array}
\end{align}
where (\ref{wenjie_25}) is a quadratic function related to the variable ${{\lambda ^i}}$.

It is worth noting that there are two different cases for the quadratic function in (\ref{wenjie_25}), including the positive value of term ${{C_2}}$ and the negative value of term ${{C_2}}$. For these two cases, terms ${{C_2}}$, ${{\lambda ^i}}$ are calculated in different ways. Specifically, the adaptive step size of each case for the A-GD algorithm can be written as
\begin{align}\label{wenjie_29}
{\lambda ^i} = \left\{ {\begin{array}{*{20}{c}}
{{{ - {C_1}} \mathord{\left/
 {\vphantom {{ - {C_1}} {\left( {2{C_2}} \right),{C_2} > 0}}} \right.
 \kern-\nulldelimiterspace} {\left( {2{C_2}} \right),{C_2} > 0}}}\\
{{{\left| {{C_1}} \right|} \mathord{\left/
 {\vphantom {{\left| {{C_1}} \right|} {\left| {{C_2}} \right|}}} \right.
 \kern-\nulldelimiterspace} {\left| {{C_2}} \right|}},\begin{array}{*{20}{c}}
{}
\end{array}{C_2} < 0}
\end{array}} \right. .
\end{align}

The detailed steps of the A-GD algorithm for solving (\ref{system_6}) are illustrated in \textbf{Algorithm 1}.

\section{Proposed Alternating Optimization Scheme}
In this section, a novel AO algorithm is proposed to decrease the heavy calculation burden experienced by Algorithm 1 and the number of variables where the multiple matrix variables involve $\left( {{{\bf{F}}_{{\rm{BB}}}},{{\bf{F}}_{{\rm{RF}}}},{{\bf{W}}_{{\rm{BB}}}},{{\bf{W}}_{{\rm{RF}}}},{\bf{\Phi }}} \right)$. Concretely, the AO algorithm contains a CBC algorithm to seek the optimal hybrid  precoders (or combiners) and a linear search algorithm to optimize the phase shift matrix.

\subsection{CBC Algorithm}
To start with, we first propose the CBC algorithm to design $\textbf{F}_{\rm RF}$, $\textbf{F}_{\rm BB}$, $\textbf{W}_{\rm RF}$ and $\textbf{W}_{\rm BB}$ with a given $\bm{\Phi}$.
However, the multiple matrix variable optimization for maximizing $R$ is intractable. As a common practice, the hybrid precoder and combiner design can be usually transformed into two separate subproblems, i.e., one precoding problem about $\left( {{{\bf{F}}_{{\rm{BB}}}},{{\bf{F}}_{{\rm{RF}}}}} \right)$, and the other combing problem about $\left( {{{\bf{W}}_{{\rm{BB}}}},{{\bf{W}}_{{\rm{RF}}}}} \right)$, \cite{yanlongfei_02},\cite{yanlongfei_03}. Thus, the precoding problem can be formulated as
\begin{equation}
\begin{aligned}
(\textbf{F}^{\rm opt}_{\rm BB},\textbf{F}^{\rm opt}_{\rm RF})=&{\mathop{\rm arg \ min}\limits_{\textbf{F}_{\rm BB},\textbf{F}_{\rm RF}}}\lVert \textbf{F}^{\rm opt}-\textbf{F}_{\rm RF}\textbf{F}_{\rm BB}\rVert_{F}^2\\
\mathrm{s.t.}\  &\textbf{F}_{\rm RF}\in\mathcal{F}_{RF},  \lVert \textbf{F}_{\rm RF}\textbf{F}_{\rm BB}\rVert^{2}_{F}=N_{s},
\end{aligned}
\label{problem_precoding}
\end{equation}
where ${{\bf{F}}^{{\rm{opt}}}} = {{\bf{V}}_1}$ is the optimal fully-digital precoding matrix \cite{introduction_33}. Similar to the precoding problem (\ref{problem_precoding}), the combining problem can be also formulated as
\begin{equation}
\begin{aligned}
(\textbf{W}^{\rm opt}_{\rm BB},\textbf{W}^{\rm opt}_{\rm RF})=&{\mathop{\rm arg \ min}\limits_{\textbf{W}_{\rm BB},\textbf{W}_{\rm RF}}}\lVert \textbf{W}^{\rm opt}-\textbf{W}_{\rm RF}\textbf{W}_{\rm BB}\rVert_{F}^2\\
\mathrm{s.t.}\  &\textbf{W}_{\rm RF}\in\mathcal{W}_{{\mathop{\rm RF}\nolimits}},
\end{aligned}
\label{problem_combining}
\end{equation}
where ${{\bf{W}}^{{\rm{opt}}}} = {{\bf{U}}_1}$ is the optimal fully-digital combining matrix. Next, the CBC algorithm is presented to solve \eqref{problem_precoding} and \eqref{problem_combining}. Since these two problems share the similar form except that the precoding problem has an additional constraint $\lVert \textbf{F}_{\rm RF}\textbf{F}_{\rm BB}\rVert^{2}_{F}=N_{s}$, we take the solution of the precoding problem as an example.

The objective of \eqref{problem_precoding} is to make $\textbf{F}_{\rm RF}\textbf{F}_{\rm BB}$ approach $\textbf{F}_{\rm opt}$. However, the main obstacle in \eqref{problem_precoding} is that each element in $\textbf{F}_{\rm RF}$ has a constant-magnitude constraint. We denote the $l$th column of $\textbf{F}_{\rm opt}$ as $\textbf{f}_{\rm opt}^{(l)}$. Interestingly, one property of $\textbf{f}_{\rm opt}^{(l)}$ under the constant-magnitude constraint is that, an arbitrary vector without constant-magnitude constraint can be expressed as the linear combination of two vectors under the constant-magnitude constraint \cite{yanlongfei_04}. Thus, the $l$th column of $\textbf{F}_{{\rm opt}}$ can be expressed as
\begin{equation}	
\textbf{f}_{\rm opt}^{(l)}=a_{l}\textbf{p}_{l}+b_{l}\textbf{q}_{l},
\label{iii_2}
\end{equation}
where $\textbf{p}_{l}$ and $\textbf{q}_{l}$ are two vectors under the constant-magnitude constraint, $a_l$ and $b_l$ are two linear combination factors. One solution of \eqref{iii_2} is that $a_l=b_l=d_{max}$, where $d_{max}$ is the largest magnitude of the elements in $\textbf{f}_{\rm opt}^{(l)}$, and then calculate $\textbf{p}_{l}$ and $\textbf{q}_{l}$. With such a feature of the constant-magnitude vector, we can set the $(2l-1)$th column and the $(2l)$th column of $\textbf{F}_{{\rm RF}}$ as $\textbf{f}_{{\rm RF}}^{(2l-1)}=\textbf{p}_l$ and $\textbf{f}_{{\rm RF}}^{(2l)}=\textbf{q}_l$. Meanwhile, the $(2l-1)$th and $(2l)$th elements of $\textbf{f}_{{\rm BB}}^{(l)}$ are set as $d_{2l-1}=a_l$ and $d_{2l}=b_l$. As a result, we have
\begin{equation}	
\setlength{\arraycolsep}{1pt}
\begin{aligned}
\begin{array}{l}
{\bf{f}}_{{\rm{opt}}}^{(l)} = {{\bf{F}}_{{\rm{RF}}}}{\bf{f}}_{{\rm{BB}}}^{(l)}
 = \left[ { \ldots ,{\bf{f}}_{{\rm{RF}}}^{(2l - 1)},{\bf{f}}_{{\rm{RF}}}^{(2l)}, \ldots } \right]{\left[ {0, \ldots ,0,{d_{2l - 1}},{d_{2l}},0, \ldots ,0} \right]^T}.
\end{array}
\end{aligned}
\label{iii}%
\end{equation}

Following (\ref{iii}), each $\textbf{f}^{(l)}_{{\rm opt}}$ can be expressed as the linear combination of two column vectors in $\textbf{F}_{{\rm RF}}$.
Recall that the dimensions of $\textbf{F}_{{\rm opt}}$ and $\textbf{F}_{{\rm RF}}$ are $N_{{\mathop{\rm BS}\nolimits}}\times N_s$ and $N_{{\mathop{\rm BS}\nolimits}}\times M_{{\mathop{\rm BS}\nolimits}}$. To transmit $N_s$ data streams, the number of RF chains $M_{{\mathop{\rm BS}\nolimits}}$ is no smaller than $N_s$. Thus, we divide the minimization of $\lVert\textbf{F}_{\rm opt}-\textbf{F}_{\rm RF}\textbf{F}_{\rm BB}\rVert_{F}^2$ into two cases according to the relation between $M_{{\mathop{\rm BS}\nolimits}}$ and $2N_s$ as follows.

\textit{\textbf{Case 1 ($M_{{\mathop{\rm BS}\nolimits}}\geq2N_s$)}}: All $\textbf{f}^{(l)}_{{\rm opt}}$ can be expressed as (\ref{iii}), and then $\textbf{F}_{{\rm opt}}$ can be represented as
\begin{equation}
\setlength{\arraycolsep}{1pt}
\textbf{F}_{{\rm opt}}=\textbf{Q}\textbf{D}=\left[\begin{array}{ccccc}
\textbf{p}_1,&\textbf{q}_1,&\ldots,&\textbf{p}_{N_s},&\textbf{q}_{N_s}
\end{array}\right]\left[\begin{array}{ccc}
\textbf{d}_1&&\\
&\ddots&\\
&&\textbf{d}_{N_s}
\end{array}\right],
\label{case_1}
\end{equation}
where $\textbf{d}_{l}=[a_l,b_l]^T$. Since $M_{{\mathop{\rm BS}\nolimits}}\geq 2N_s$, we can make the first $2N_s$ columns of $\textbf{F}_{{\rm RF}}$ as $\textbf{Q}$ and the first $2N_s$ rows of $\textbf{F}_{{\rm BB}}$ as $\textbf{D}$. Meanwhile, the remaining $(M_{{\mathop{\rm BS}\nolimits}}-2N_s)$ rows of $\textbf{F}_{{\rm BB}}$ can be set zero, and the remaining $(M_{{\mathop{\rm BS}\nolimits}}-2N_s)$ columns of $\textbf{F}_{{\rm RF}}$ can be set arbitrarily. Consequently, $\textbf{F}_{{\rm opt}}$ can be exactly expressed as $\textbf{F}_{{\rm RF}}\textbf{F}_{{\rm BB}}$, i.e.,$\left\| {{{\bf{F}}_{{\rm{opt}}}} - {{\bf{F}}_{{\rm{RF}}}}{{\bf{F}}_{{\rm{BB}}}}} \right\|_F^2 = 0$, which indicates that the optimal $\textbf{F}_{{\rm RF}}$ and $\textbf{F}_{{\rm BB}}$ are obtained accordingly.

\textit{\textbf{Case 2 ($N_s\leq M_{{\mathop{\rm BS}\nolimits}}<2N_s$)}}: Since $M_{{\mathop{\rm BS}\nolimits}}<2N_s$, the columns of $\textbf{F}_{{\rm RF}}$ and the rows of $\textbf{F}_{{\rm BB}}$ are insufficient to be set as $\textbf{Q}$ and $\textbf{D}$. Fortunately, according to \cite{yanlongfei_05}, we note that $||{\bf{f}}_{{\rm{opt}}}^{(l)}|{|_\infty } \le \sqrt {{{{L_p}} \mathord{\left/ {\vphantom {{{L_p}} {{N_{{\rm{BS}}}}}}} \right. \kern-\nulldelimiterspace} {{N_{{\rm{BS}}}}}}} $, where $L_p$ is the number of multipath of the THz channel. This property implies that the amplitude differences of the elements in $\textbf{f}_{{\rm opt}}^{(l)}$ are smaller than $\sqrt {{{{L_p}} \mathord{\left/ {\vphantom {{{L_p}} {{N_{{\rm{BS}}}}}}} \right. \kern-\nulldelimiterspace} {{N_{{\rm{BS}}}}}}}$. Due to the sparse nature of the THz channel, $L_p$ is usually small \cite{introduction_09}.
Considering the deployment of ultra-massive antennas at THz band, e.g., ${{N_{{\rm{BS}}}}}=$512 antennas, $\sqrt {{{{L_p}} \mathord{\left/ {\vphantom {{{L_p}} {{N_{{\rm{BS}}}}}}} \right. \kern-\nulldelimiterspace} {{N_{{\rm{BS}}}}}}}$ is far less than $1$. As a result, the amplitude of each element in $||{\bf{f}}_{{\rm{opt}}}^{(l)}{{|{|_1}} \mathord{\left/ {\vphantom {{|{|_1}} {{N_{{\rm{BS}}}}}}} \right. \kern-\nulldelimiterspace} {{N_{{\rm{BS}}}}}}$, which guarantees that $\textbf{f}_{{\rm opt}}^{(l)}$ can be approximated as
\begin{equation}
\textbf{f}_{{\rm opt}}^{(l)}\approx\textbf{F}\!_{{\rm RF}}\textbf{f}_{{\rm BB}}^{(l)}\!= [...,\textbf{f}_{{\rm RF}}^{(l)},...][0,...,0,d_l,0,...,0]^T,
\label{ggg}
\end{equation}
where $d_l$ is the $l$th element of $\textbf{f}_{{\rm BB}}^{(l)}$. In addition, the optimal $d_l$ and $\textbf{f}_{\rm RF}^{(l)}$ for $\textbf{f}_{\rm opt}^{(l)}$ are known as
\begin{align}
&d_l=\frac{\lVert\textbf{f}_{{\rm opt}}^{(l)}\rVert_1}{N_{{\mathop{\rm BS}\nolimits}}},\\
\textbf{f}_{{\rm RF}}^{(l)}=[e^{j*\frac{f_{1}}{\lvert f_{1}\rvert}}&,...,e^{j*\frac{f_{ q}}{\lvert f_{q}\rvert}},...,e^{j*\frac{f_{ N_{{\mathop{\rm BS}\nolimits}}}}{\lvert f_{N_{{\mathop{\rm BS}\nolimits}}}\rvert}}]^T,
\end{align}
where ${f}_{q}$ is the $q$th element of $\textbf{f}_{{\rm opt}}^{(l)}$. The phase information of $\textbf{f}_{\rm opt}^{(l)}$ is contained in $\textbf{f}_{\rm RF}^{(l)}$, while the approximation error comes from the amplitude differences between the element in $\textbf{f}_{\rm opt}^{(l)}$ and $||{\bf{f}}_{{\rm{opt}}}^{(l)}{{|{|_1}} \mathord{\left/ {\vphantom {{|{|_1}} {{N_{{\rm{BS}}}}}}} \right. \kern-\nulldelimiterspace} {{N_{{\rm{BS}}}}}}$, which is small as we have analyzed before.

Currently, there are two approaches to express the vector $\textbf{f}_{{\rm opt}}^{(l)}$. One is to utilize two columns in $\textbf{F}_{{\rm RF}}$ to exactly express $\textbf{f}_{{\rm opt}}^{(l)}$ as shown in (\ref{iii}), and the other is to employ one column in $\textbf{F}_{{\rm RF}}$ to approximate $\textbf{f}_{{\rm opt}}^{(l)}$ as shown in (\ref{ggg}). Since the first method suffers the less performance degradation, we aim to express the columns of $\textbf{f}_{{\rm opt}}^{(l)}$ by (\ref{iii}). Under the case of $M_{{\mathop{\rm BS}\nolimits}}<2N_s$, we at most select $(M_{{\mathop{\rm BS}\nolimits}}-N_s)$ columns of $\textbf{f}_{{\rm opt}}^{(l)}$ that are expressed by (\ref{iii}), while the remaining $(2N_s-M_{{\mathop{\rm BS}\nolimits}})$ columns need to be approximated by (\ref{ggg}). One further question is how to classify these $(M_{{\mathop{\rm BS}\nolimits}}-N_s)$ columns and $(2N_s-M_{{\mathop{\rm BS}\nolimits}})$ columns. Noting that the performance penalty is only caused by the approximation error in (\ref{ggg}), we adopt the variance $\sigma^2_l$ to quantify the amplitude difference of each element in $\textbf{f}_{{\rm opt}}^{(l)}$, and the $\sigma^2_l$ can be defined as
\begin{equation}
\sigma _l^2 \buildrel \Delta \over = \sum\limits_{q = 1}^{{N_{{\rm{BS}}}}} {{{\left( {{f_q} - \frac{{{{\left\| {{\bf{f}}_{{\rm{opt}}}^{(l)}} \right\|}_1}}}{{{N_{{\rm{BS}}}}}}} \right)}^2}},
\label{mmm}
\end{equation}
where $f_q$ is the amplitude of the $q$th element in $\textbf{f}_{{\rm opt}}^{(l)}$. Based on the classification criteria,  we select $(2N_s-M_{{\mathop{\rm BS}\nolimits}})$ columns with the least variance $\sigma^2_l$ from $\textbf{f}^{(l)}_{{\rm opt}}$ to be approximated as (\ref{ggg}), and then the remaining $(M_{{\mathop{\rm BS}\nolimits}}-N_s)$ columns are calculated as (\ref{iii}). The important steps of the CBC algorithm for the precoding problem \eqref{problem_precoding} are presented in \textbf{Algorithm 2}. Similarly, the combining problem \eqref{problem_combining} can also be handled by the CBC algorithm without the normalization step $\textbf{F}_{\rm BB}\leftarrow\frac{\sqrt{N_s}}{\ \ \left\lVert{\textbf{F}}_{\rm RF}\textbf{F}_{\rm BB}\right\rVert_{F}}\textbf{F}_{\rm BB}$.

\begin{algorithm}[!t]
	\caption{CBC Algorithm}
	\begin{algorithmic}[1]
		\REQUIRE ${{{\bf{H}}_{\rm{1}}}}$, ${{{\bf{H}}_{\rm{2}}}}$, ${M_{{\rm{BS}}}}$, ${{N_s}}$, $\textbf{F}^{{\rm opt}}$, ${{\bf{\Phi }}} = \bar \mu {{\bf{I}}_{{N_{{\rm{RIS}}}} \times {N_{{\rm{RIS}}}}}}$
        \STATE \textbf{\emph{Case 1}:} Compute $\textbf{F}_{{\rm RF}}$ and $\textbf{F}_{{\rm BB}}$ as (\ref{case_1});
        \STATE \textbf{\emph{Case 2}:}
        \STATE ${{}}$ ${{}}$ Compute $\sigma^2_l$ of $\textbf{f}^{(l)}_{{\rm opt}}$,
        \STATE ${{}}$ ${{}}$ Use the \textit{classification criteria} (\ref{mmm}),
        \STATE ${{}}$ ${{}}$ \textbf{for} $l=1:N_s$
        \STATE ${{}}$ ${{}}$ ${{}}$ ${{}}$ Calculate $\textbf{f}^{(l)}_{{\rm RF}}$ and $\textbf{f}^{(l)}_{{\rm BB}}$ as (\ref{iii}) or (\ref{ggg}),
        \STATE ${{}}$ ${{}}$ \textbf{end for}
        \STATE ${{}}$ ${{}}$  Normalize $\textbf{F}_{\rm BB}$ as $\textbf{F}_{\rm BB}\leftarrow\frac{\sqrt{N_s}}{\ \ \left\lVert{\textbf{F}}_{\rm RF}\textbf{F}_{\rm BB}\right\rVert_{F}}\textbf{F}_{\rm BB}$.
        \ENSURE $\textbf{F}_{{\rm RF}}$, $\textbf{F}_{{\rm BB}}$
	\end{algorithmic}
\end{algorithm}

\subsection{Linear Search Algorithm}
With the given precoders and combiners solved by CBC algorithm, the linear search algorithm aims to optimize ${{N_{{\rm{RIS}}}}}$ reflection coefficient variables ${\left\{ {{\phi _n} = {\bar \mu} {e^{j{\varphi _n}}}} \right\}_{n = 1}^{{N_{{\rm{RIS}}}}}}$ by following the one-the-rest criterion. Concretely, this criterion indicates that only one phase shift variable can be selected from a finite phase shift set ${{\cal F}}$ while the rest $({{N_{{\rm{RIS}}}} - 1})$ phase shift variables remain fixed. Remarkably, our proposed linear search algorithm endures the linear complexity growth with the increasing number of RIS elements.

According to (\ref{system_7}), the data rate of the RIS-enabled THz MIMO system can be formulated as
\begin{align}\label{mage_1}
\begin{array}{l}
R = {\log _2}\left| {{{\bf{I}}_{{N_s}}} + \frac{\rho }{{{\delta ^2}{N_s}}}{{\left( {{{\bf{W}}^H}{\bf{W}}} \right)}^{ - 1}}{{\bf{W}}^H}\left( {{{\bf{H}}_2}{\bf{\Phi }}{{\bf{H}}_1}} \right)} { {\bf{F}}{{\bf{F}}^H}{{\left( {{{\bf{H}}_2}{\bf{\Phi }}{{\bf{H}}_1}} \right)}^H}{\bf{W}}} \right|\\
\;\;\;\; \mathop  \ge \limits^{(e)} {\log _2}\left( {\left| {{{\bf{I}}_{{N_s}}}} \right| + \left| {\frac{\rho }{{{\delta ^2}{N_s}}}{{\left( {{{\bf{W}}^H}{\bf{W}}} \right)}^{ - 1}}\left( {{{\bf{W}}^H}{{\bf{H}}_2}{\bf{\Phi }}{{\bf{H}}_1}{\bf{F}}} \right)} \right.} {\left. { {{\left( {{{\bf{W}}^H}{{\bf{H}}_2}{\bf{\Phi }}{{\bf{H}}_1}{\bf{F}}} \right)}^H}} \right|} \right)\\
\;\;\;\; \mathop  = \limits^{(f)} {\log _2}\left( {\left| {{{\bf{I}}_{{N_s}}}} \right| + \left| {{{\left( {{{\bf{W}}^H}{\bf{W}}} \right)}^{ - 1}}} \right|\left| {\frac{\rho }{{{\delta ^2}{N_s}}}\left( {{{\bf{W}}^H}{{\bf{H}}_2}{\bf{\Phi }}{{\bf{H}}_1}{\bf{F}}} \right)} \right.} {\left. {{{\left( {{{\bf{W}}^H}{{\bf{H}}_2}{\bf{\Phi }}{{\bf{H}}_1}{\bf{F}}} \right)}^H}} \right|} \right),
\end{array}
\end{align}
where the definition $(e)$ satisfies ${\left| {{\bf{B}} + {\bf{C}}} \right| \ge \left| {\bf{B}} \right| + \left| {\bf{C}} \right|}$ for any positive semidefinite matrices ${{\bf{B}}}$ and ${{\bf{C}}}$, and the definition $(f)$ holds ${\left| {{\bf{BC}}} \right| = \left| {\bf{B}} \right|\left| {\bf{C}} \right|}$ for any square matrices ${{\bf{B}}}$ and ${{\bf{C}}}$. Since both precoding matrix ${{\bf{F}}}$ and combining matrix ${{\bf{W}}}$ are constant matrices, the achievable rate maximization problem is further simplified as
\begin{align}\label{mage_2}
\begin{array}{l}
{{\bf{\Phi }}^{{\rm{opt }}}} = \mathop {\arg \max }\limits_{\bf{\Phi }} \;\tilde R\\
s.t.\;\;{\rm{ }}{\varphi _n} \in {{\cal F}},\;{\varphi _{\max }} = {306.82^{\rm{o}}},\;\forall n = 1, \ldots ,{N_{{\rm{RIS}}}},\\
\;\;\;\;\;\;\;\left| {{\phi _n}} \right| = \left| {\bar \mu {e^{j{\varphi _n}}}} \right| = 0.8,\;\forall n = 1, \ldots ,{N_{{\rm{RIS}}}},
\end{array}
\end{align}
where we define ${\tilde R}$ as
\begin{align}\label{mage_3}
\begin{array}{l}
\tilde R = {\log _2}\left| {\frac{\rho }{{{\delta ^2}{N_s}}}\left( {{{\bf{W}}^H}{{\bf{H}}_2}{\bf{\Phi }}{{\bf{H}}_1}{\bf{F}}} \right)} { {{\left( {{{\bf{W}}^H}{{\bf{H}}_2}{\bf{\Phi }}{{\bf{H}}_1}{\bf{F}}} \right)}^H}} \right|,
\end{array}
\end{align}

Here we define ${{{\bf{\bar H}}_1} = {{\bf{H}}_1}{\bf{F}} = {\left[ {{{{\bf{\bar h}}}_{1,1}}, \cdots ,{{{\bf{\bar h}}}_{1,{N_{{\rm{RIS}}}}}}} \right]^H}}$ and ${{{\bf{\bar H}}_2} = {{\bf{W}}^H}{{\bf{H}}_2} = \left[ {{{{\bf{\bar h}}}_{2,1}}, \cdots ,{{{\bf{\bar h}}}_{2,{N_{{\rm{RIS}}}}}}} \right]}$, where ${{{\bf{\bar h}}_{1,n}} \in {{\mathbb C}^{{N_s} \times 1}}}$ and ${{{\bf{\bar h}}_{2,n}} \in {{\mathbb C}^{{N_s} \times 1}}}$. With the defined form, ${{{\bf{\bar H}}_{\rm{e}}}}$ can be expressed as
\begin{align}\label{mage_4}
{{\bf{\bar H}}_{\rm{e}}} = \left( {{{{\bf{\bar H}}}_2}{\bf{\Phi }}{{{\bf{\bar H}}}_1}} \right) = \sum\limits_{n = 1}^{{N_{{\rm{RIS}}}}} {{\phi _n}{{{\bf{\bar h}}}_{2,n}}{\bf{\bar h}}_{1,n}^H},
\end{align}

By resorting to (\ref{mage_4}), we are able to rewrite (\ref{mage_3}) as
\begin{align}\label{mage_5}
\begin{array}{l}
\tilde R = {\log _2}\left| {\frac{\rho }{{{\delta ^2}{N_s}}}\left( {{{{\bf{\bar H}}}_2}{\bf{\Phi }}{{{\bf{\bar H}}}_1}} \right){{\left( {{{{\bf{\bar H}}}_2}{\bf{\Phi }}{{{\bf{\bar H}}}_1}} \right)}^H}} \right| \\ \;\;\;\;= {\log _2}\left| {\frac{\rho }{{{\delta ^2}{N_s}}}\left( {\sum\limits_{n = 1}^{{N_{{\rm{RIS}}}}} {{\phi _n}{{{\bf{\bar h}}}_{2,n}}{\bf{\bar h}}_{1,n}^H} } \right)} {{{\left( {\sum\limits_{n = 1}^{{N_{{\rm{RIS}}}}} {{\phi _n}{{{\bf{\bar h}}}_{2,n}}{\bf{\bar h}}_{1,n}^H} } \right)}^H}} \right|\\
\begin{array}{*{20}{c}}
{}
\end{array} = {\log _2}\left| {\frac{\rho }{{{\delta ^2}{N_s}}}\left( {\sum\limits_{n = 1}^{{N_{{\rm{RIS}}}}} {{\phi _n}{{{\bf{\bar h}}}_{2,n}}{\bf{\bar h}}_{1,n}^H} } \right)} {\left( {\sum\limits_{n = 1}^{{N_{{\rm{RIS}}}}} {\phi _n^ * {{{\bf{\bar h}}}_{1,n}}{\bf{\bar h}}_{2,n}^H} } \right)} \right|,
\end{array}
\end{align}

To solve the optimization problem with ${{N_{{\rm{RIS}}}}}$ reflection coefficient variables, the linear search algorithm is investigated under the criterion of one-the-rest, which guarantees to optimize only one variable at a time. Concretely, when we optimize ${{\phi _n}}$, the rest variables ${\left\{ {{\phi _i}} \right\}_{i = 1,i \ne n}^{{N_{{\rm{RIS}}}}}}$ are treated as constants during the optimization process. On the basic of this criterion, the objective function in terms of only one variable ${{\phi _n}}$ can be rewritten as
\begin{align}\label{mage_6}
\begin{array}{l}
\tilde R\left( {{\phi _n}} \right) = {\log _2}\left| {\frac{\rho }{{{\delta ^2}{N_s}}}\left( {{\phi _n}{{{\bf{\bar h}}}_{2,n}}{\bf{\bar h}}_{1,n}^H + \sum\limits_{i = 1,i \ne n}^{{N_{{\rm{RIS}}}}} {{\phi _i}{{{\bf{\bar h}}}_{2,i}}{\bf{\bar h}}_{1,i}^H} } \right)} {\left( {\phi _n^ * {{{\bf{\bar h}}}_{1,n}}{\bf{\bar h}}_{2,n}^H + \sum\limits_{j = 1,j \ne n}^{{N_{{\rm{RIS}}}}} {\phi _j^ * {{{\bf{\bar h}}}_{1,j}}{\bf{\bar h}}_{2,j}^H} } \right)} \right|\\
\begin{array}{*{20}{c}}
{}&{}
\end{array} \;\;\;\; = {\log _2}\left| {\frac{\rho }{{{\delta ^2}{N_s}}}\left( {{{\bf{P}}_n} + {\phi _n}{{\bf{Q}}_n} + \phi _n^ * {\bf{Q}}_n^H} \right)} \right|,
\end{array}
\end{align}
where
\begin{align}\label{mage_7}
\begin{array}{l}
{{\bf{P}}_n} = {{\bar \mu} ^2}{{{\bf{\bar h}}}_{2,n}}{\bf{\bar h}}_{1,n}^H{{{\bf{\bar h}}}_{1,n}}{\bf{\bar h}}_{2,n}^H + \left( {\sum\limits_{i = 1,i \ne n}^{{N_{{\rm{RIS}}}}} {{\phi _i}{{{\bf{\bar h}}}_{2,i}}{\bf{\bar h}}_{1,i}^H} } \right)\left( {\sum\limits_{j = 1,j \ne n}^{{N_{{\rm{RIS}}}}} {\phi _j^ * {{{\bf{\bar h}}}_{1,j}}{\bf{\bar h}}_{2,j}^H} } \right)\\
{{\bf{Q}}_n} = {{{\bf{\bar h}}}_{2,n}}{\bf{\bar h}}_{1,n}^H\left( {\sum\limits_{j = 1,j \ne n}^{{N_{{\rm{RIS}}}}} {\phi _j^ * {{{\bf{\bar h}}}_{1,j}}{\bf{\bar h}}_{2,j}^H} } \right).
\end{array}
\end{align}

Through the above discussion, we can see that both ${{{\bf{P}}_n}}$ and ${{{\bf{Q}}_n}}$ are independent of ${{\phi _n}}$. Under this condition, the phase shift optimization of ${N_{{\rm{RIS}}}}$ reflecting elements in (\ref{mage_2}) can be converted into the problem (\ref{mage_6}) that just needs to optimize the phase shift of each reflecting element. Therefore, with respect to ${{\phi _n}}$, the optimization objective can be formulated as
\begin{align}\label{mage_8}
\begin{array}{l}
\phi _n^{{\rm{opt}}} = \mathop {\arg \;\max }\limits_{{\phi _n}} \;{\log _2}\left| {\frac{\rho }{{{\delta ^2}{N_s}}}\left( {{{\bf{P}}_n} + {\phi _n}{{\bf{Q}}_n} + \phi _n^ * {\bf{Q}}_n^H} \right)} \right|\\
s.t.\;\;\;\;\;{\phi _n} = \bar \mu {e^{j{\varphi _n}}},{\varphi _n} \in {\cal F}.
\end{array}
\end{align}

\begin{algorithm}[!t]
	\caption{Linear Search Algorithm}
	\begin{algorithmic}[1]
		\REQUIRE ${{{\bf{H}}_1}}$, ${{{\bf{H}}_2}}$, ${{\bf{F}}}$, ${{\bf{W}}}$, ${\rho }$, ${{N_s}}$, ${{\delta ^2}}$, ${b}$, ${{\varphi _{\max }}}$, ${\cal F}$,
		\STATE Initialize the phase shift matrix ${{\bf{\Phi }} = \;\left( {{\phi _1},{\phi _2}, \cdots ,{\phi _{{N_{{\rm{RIS}}}}}}} \right)}$,
        \STATE \textbf{Iterative Process:}
        \STATE \textbf{for} ${n = 1:{N_{{\rm{RIS}}}}}$ \textbf{do}
        \STATE ${{}}$ ${{}}$ ${{}}$\textbf{for all} ${{\varphi _n} \in {\cal F}}$ \textbf{do}
        \STATE ${{}}$ ${{}}$ ${{}}$ ${{}}$ Calculate ${{{\bf{P}}_n}}$ and ${{{\bf{Q}}_n}}$ based on (\ref{mage_7}),
        \STATE ${{}}$ ${{}}$ ${{}}$ ${{}}$ Obtain the optimal $\phi _n^{{\rm{opt}}} = \bar \mu {e^{j\varphi _n^{{\rm{opt}}}}}$ according to (\ref{mage_8}),
        \STATE ${{}}$ ${{}}$ ${{}}$ ${{}}$ Update the ${n}$th entry of ${{\bf{\Phi }}}$: ${{\phi _n} = \phi _n^{{\rm{opt}}}}$,
        \STATE ${{}}$ ${{}}$ ${{}}$\textbf{end for}
        \STATE \textbf{end for}
        \ENSURE ${{{\bf{\Phi }}^{{\rm{opt}}}} = \;diag \left( {\phi _1^{{\rm{opt}}},\phi _2^{{\rm{opt}}}, \cdots ,\phi _{{N_{{\rm{RIS}}}}}^{{\rm{opt}}}} \right)}$, $R$
	\end{algorithmic}
\end{algorithm}

Nevertheless, due to the discrete phase shift ${{\varphi _n}}$, this single variable optimization problem (\ref{mage_8}) is still non-convex. It should be noted that the discrete phase shift ${{\varphi _n}}$ is selected from a predefined phase shift set ${{\cal F}}$ and is finite. Hence, the non-convex optimization problem can be considered as a phase shift search problem. The detailed process of our proposed linear search algorithm is illustrated in \textbf{Algorithm 3}.

%
%
%
%
%
\subsection{Complexity Analysis}
In this subsection, the complexity of our proposed A-GD and AO algorithms is analyzed in detail, and the complexity represents the total number of the complex multiplications. We first analyze the complexity of the AO algorithm that is the combination of the CBC algorithm and the linear search algorithm. In terms of the CBC algorithm, the complexity for \emph{Case 1} is $\mathcal{O}(N_{{{\mathop{\rm BS}\nolimits} }}N_{s})$ while the complexity for the \emph{Case 2} is $\mathcal{O}\left( {{N_{{\rm{BS}}}}{M_{{\rm{BS}}}}{N_s} + 2{N_{{\rm{BS}}}}{N_s} + {N_{{\rm{BS}}}}} \right)$. Hence, the maximum complexity for precoding problem equals to the complexity of \emph{Case 2}. Similarly, for combining problem, the complexity is $\mathcal{O}\left( {{N_{{\rm{MS}}}}{N_s} + {N_{{\rm{MS}}}}} \right)$. To sum up, the overall complexity of the CBC algorithm can be expressed as
\begin{align}\label{complexity_1}
\mathcal{O}_{\rm{CBC}}\left( \begin{array}{l}
{N_{{\rm{BS}}}}{M_{{\rm{BS}}}}{N_s} + 2{N_{{\rm{BS}}}}{N_s} + {N_{{\rm{BS}}}}\\
 + {N_{{\rm{MS}}}}{N_s} + {N_{{\rm{MS}}}}
\end{array} \right).
\end{align}

For the linear search algorithm, there are total ${{\cal O}\left( {{N_{{\mathop{\rm RIS}\nolimits} }}{2^b}} \right)}$ possible phase shift combinations, and the complexity of calculating each ${\phi _n}$ is $\mathcal{O}\left( {2{N_{{\rm{RIS}}}}N_s^2 + N_s^3} \right)$. Thus, the complexity of the linear search algorithm can be written as
\begin{align}\label{complexity_2}
\mathcal{O}_{\rm{LS}}\left( {{2^{b + 1}}N_{{\rm{RIS}}}^2N_s^2 + {2^b}{N_{{\rm{RIS}}}}N_s^3} \right).
\end{align}

Based on (\ref{complexity_1}) and (\ref{complexity_2}), the complexity of the AO algorithm with ${I_o}$ iterations is written as
\begin{align}\label{complexity_3}
{{{\cal O}}_{{\rm{AO}}}}\left[ {{I_o}\left( \begin{array}{l}
{N_{{\rm{BS}}}}{M_{{\rm{BS}}}}{N_s} + 2{N_{{\rm{BS}}}}{N_s} + {N_{{\rm{BS}}}}
 + {N_{{\rm{MS}}}}{N_s} + {N_{{\rm{MS}}}}\\
 + {2^{b + 1}}N_{{\rm{RIS}}}^2N_s^2 + {2^b}{N_{{\rm{RIS}}}}N_s^3
\end{array} \right)} \right].
\end{align}

With respect to the A-GD algorithm, the main complexity during the calculation process is presented as follows. Firstly, the complexity of initializing ${{\bf{A}} =  - {\left( {{\bf{H}}_1^T \otimes {{\bf{H}}_2}} \right)^H}\left( {{\bf{H}}_1^T \otimes {{\bf{H}}_2}} \right)}$ is  ${{\cal O}\left( {2N_{{\mathop{\rm BS}\nolimits} }^3N_{{\mathop{\rm MS}\nolimits} }^3N_{{\mathop{\rm RIS}\nolimits} }^6} \right)}$. Then, the complexity of computing the gradient term ${{\nabla _{\bf{\varphi }}}{\mathop{f}\nolimits} \left( {\bf{\varphi }} \right)}$ is ${{\cal O}\left( {N_{{\mathop{\rm RIS}\nolimits} }^2} \right)}$. Next, the A-GD algorithm needs to calculate the adaptive step size for each iteration process and the complexity of calculating ${\lambda ^i}$ is ${{\cal O}}\left( {{{3N_{{\rm{RIS}}}^2} \mathord{\left/ {\vphantom {{3N_{{\rm{RIS}}}^2} 2}} \right. \kern-\nulldelimiterspace} 2}} \right)$. Thus, the overall complexity of the A-GD algorithm with the maximum number of iterations ${I_a}$ can be written as
\begin{align}\label{complexity_5}
{{{\cal O}}_{{\rm{A-GD}}}}\left( {\frac{5}{2}{I_a}N_{{\rm{RIS}}}^2 + 2N_{{\rm{BS}}}^3N_{{\rm{MS}}}^3N_{{\rm{RIS}}}^6} \right).
\end{align}

In order to determine the effectiveness of our proposed AO and A-GD algorithms, we need to regard the computational complexity of the C-GD algorithm in \cite{yanlongfei_01} as a benchmark metric. In light of \cite{yanlongfei_01}, the complexity formulation of the C-GD algorithm with maximum number of iterations ${I_c}$ can be expressed as
\begin{align}\label{complexity_4}
{{{\cal O}}_{{\rm{C-GD}}}}\left( {{I_c}N_{{\rm{RIS}}}^2 + 2N_{{\rm{BS}}}^3N_{{\rm{MS}}}^3N_{{\rm{RIS}}}^6} \right).
\end{align}

\section{Simulation Results}


In this section, simulation results are provided to examine the effectiveness of the considered algorithms for the downlink RIS-enabled THz MIMO system, including A-GD algorithm, AO algorithm, random phase algorithm, and C-GD algorithm \cite{yanlongfei_01}. In addition, to demonstrate the performance improvement brought by the RIS, the conventional THz MIMO system without the RIS is also simulated under the same simulation settings as the RIS-enabled THz system. The working frequency of each RIS element is at ${f = 1.6}$ THz, the maximum phase response is ${{\varphi _{\max }} = {306.82^{\rm{o}}}}$, and the average reflecting amplitude is $\bar \mu  = 0.8$, respectively. In addition, the distances of BS-RIS, RIS-MS and BS-MS are set as ${{\bar r_0} = 10\;m}$, ${{\tilde r_0} = 20\;m}$, and ${{r_0} = 25\;m}$ respectively, which can commendably meet the indoor communication scenarios. The LoS path of BS-MS link is blocked by the obstacle, and thus the NLoS paths are assisted by the RIS. Considering the sparse nature of THz channel, we assume that the BS-RIS channel ${{{\bf{H}}_1}}$ contains ${{L_1} = 3}$ propagation paths, such as one LoS path and two NLoS paths. More specifically, the complex gain of LoS path is generated based on (\ref{system_5}) and the complex gains of NLoS paths are computed by (\ref{system_6}). The molecular absorbing coefficient and the reflection coefficient of ceramic tile are set as ${\kappa (f) = 0.2}$ and ${\xi (f) = {10^{ - 6}}}$ \cite{system_02}. Similarly, the parameter settings of ${{{\bf{H}}_2}}$ are consistent with the channel ${{{\bf{H}}_1}}$. The numbers of the RF chains at the BS and the MS are set as ${{M_{{\rm{BS}}}}}=6$ and ${{M_{{\rm{MS}}}}}=4$. To combat the serious attenuation, the antenna gains are configured as  ${{G_t} = {G_r} = 55}$ dBi \cite{yanlongfei_06}.  Here we define the signal-to-noise ratio (SNR) as ${{\rm{SNR}} = \rho /{\sigma ^2}}$, and all simulation results are averaged over 1000 random channel realizations.

\begin{figure*}[!t]
\centering
\includegraphics[width=15cm]{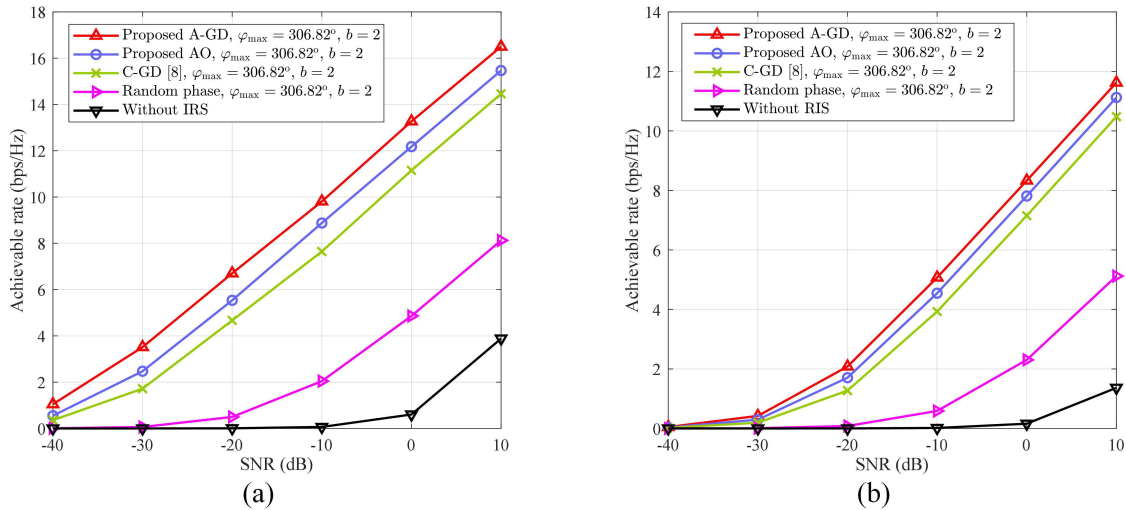}
\caption{Achievable rate comparisons of considered algorithms versus SNR: (a) ${{N_{{\rm{BS}}}}{\rm{ = 512}}}$, ${{N_{{\rm{RIS}}}}{\rm{ = 128}}}$, ${{N_{{\rm{MS}}}}{\rm{ = 32}}}$; (b) ${{N_{{\rm{BS}}}}{\rm{ = 128}}}$, ${{N_{{\rm{RIS}}}}{\rm{ = 64}}}$, ${{N_{{\rm{MS}}}}{\rm{ = 16}}}$.}\label{mage01}
\end{figure*}

Fig. \,\ref{mage01} depicts the achievable rate performance versus diverse SNR values in the THz MIMO system with different parameter settings. In both Fig. \,\ref{mage01} (a) and Fig. \,\ref{mage01} (b), the achievable rate of the RIS-enabled THz system greatly outstrips the conventional THz system without RIS that hardly meet the future communication requirements. Numerically, as shown in Fig. \,\ref{mage01} (a), the performance gap between the random phase scheme and the THz system without RIS is about 4.24 bps/Hz under the condition of SNR=10 dB. Meanwhile, Fig. \,\ref{mage01} (a) also indicates that the achievable rates of A-GD algorithm, AO algorithm and C-GD algorithm are around 8.4 bps/Hz, 7.35 bps/Hz and 6.34 bps/Hz higher than the random phase scheme, respectively. More importantly, both of our proposed A-GD algorithm and AO algorithm are superior to the C-GD algorithm that has been proposed in \cite{yanlongfei_01}, which demonstrates that our proposed optimization algorithms can be employed to further enhance the achievable rate performance for the RIS-empowered THz MIMO systems.

Fig. 6 investigates the achievable rate performance with the increasing number of the maximum phase response ${{\varphi _{\max }}}$, which aims to validate the effectiveness of the proposed hardware architecture mentioned in Section II. The parameters of the THz MIMO system are set as ${{N_{{\rm{BS}}}}{\rm{ = 512}}}$, ${{N_{{\rm{RIS}}}}{\rm{ = 128}}}$, ${{N_{{\rm{MS}}}}{\rm{ = 32}}}$. From Fig. 6 we can note that the achievable rates of the considered RIS-enabled schemes improve firstly and then converge to a fixed value. Remarkably, the achievable rates of our proposed A-GD and AO algorithms with ${{\varphi _{\max }} = {306.82^{\rm{o}}}}$ have already converged, and possess the same performance as the ideal case with ${{\varphi _{{\rm{ideal}}}} = {360^{\rm{o}}}}$. Compared with low phase response ${{\varphi _{\max }} = {60^{\rm{o}}}}$, the considered A-GD algorithm, C-GD algorithm, AO algorithm and the random phase scheme with ${{\varphi _{\max }} = {306.82^{\rm{o}}}}$ are able to realize the achievable rate enhancement of 3.59 bps/Hz, 3.50 bps/Hz, 2.79 bps/Hz and 2.68 bps/Hz, respectively. Hence, the numerical results verify the efficient hardware structure of the graphene-based RIS.

Fig. 7 provides the achievable rate performance versus the different number of bit quantization $b$ for the THz system with ${{N_{{\rm{BS}}}}{\rm{ = 512}}}$, ${{N_{{\rm{RIS}}}}{\rm{ = 128}}}$, ${{N_{{\rm{MS}}}}{\rm{ = 32}}}$. Fig. 7 indicates that the achievable rates of the random phase scheme and the conventional THz system without RIS are insensitive to the number of bit quantization. Instead, the achievable rates of our proposed A-GD and AO algorithms are evidently affected by $b$, and meanwhile surpass the C-GD algorithm. In the case of ${b = 1}$, our developed A-GD and AO algorithms suffer from the obvious performance degradation due to the limited quantization precision of RIS elements. Specifically, in contrast with ${b = 2}$, our proposed A-GD algorithm and AO algorithm with ${b = 1}$ endure about 0.93 bps/Hz and 0.95 bps/Hz performance penalty, respectively. Intriguingly, the achievable rates of our proposed algorithms with  ${b = 2}$ almost yields the similar performance compared with ${b \ge 3}$, indicating that ${b = 2}$ is sufficient to quantize the discrete phase shifts of the RIS elements in practice.

\makeatletter
\renewcommand{\@thesubfigure}{\hskip\subfiglabelskip}
\makeatother
\begin{figure} \centering
\subfigure[Fig. 6: Achievable rate comparisons of considered algorithms versus ${{\varphi _{\max }}}$.] {
 \label{mage02}
\includegraphics[width=0.43\columnwidth]{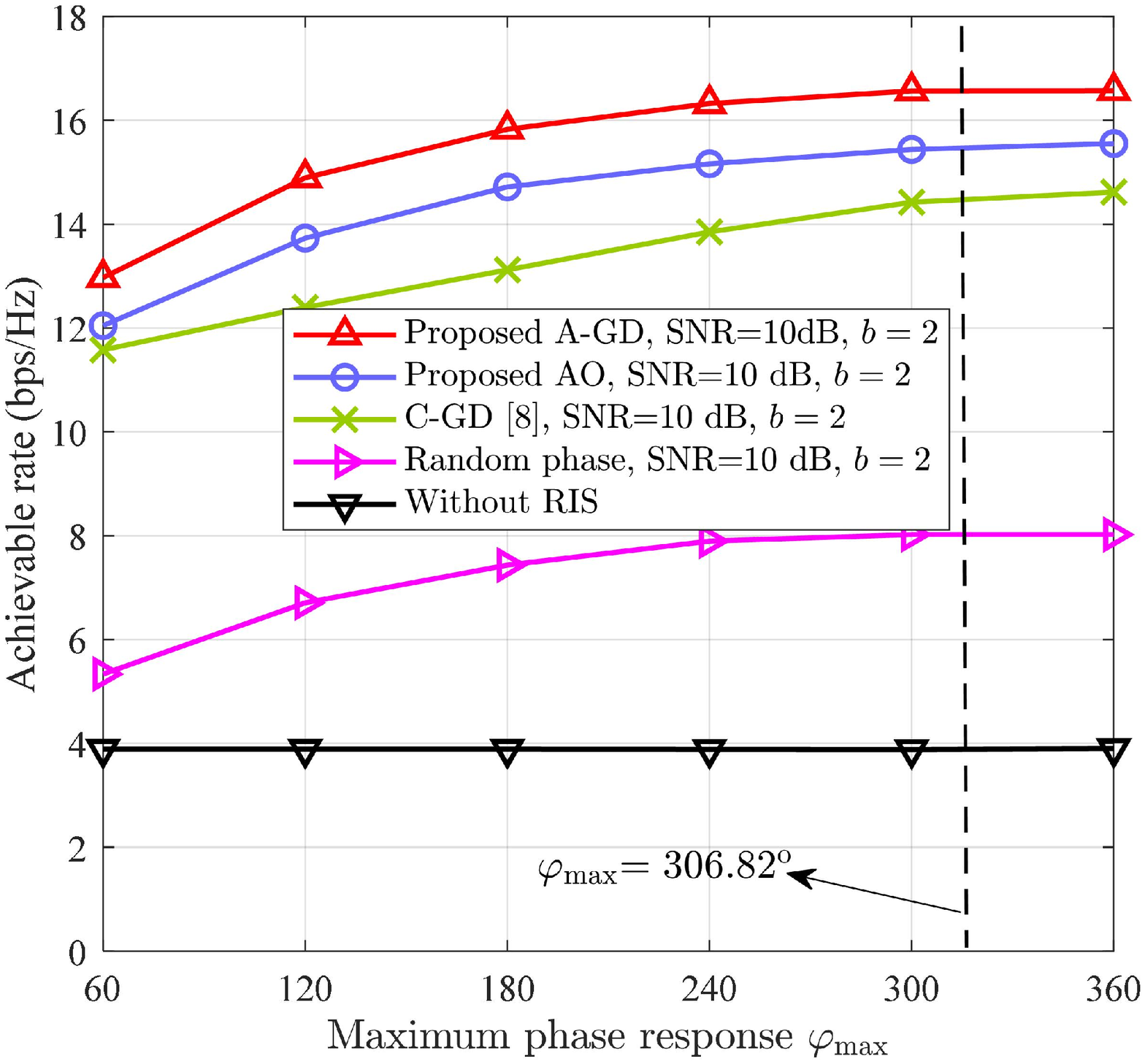}
} \;\;\;\;\;\;\;\;\;\;\;\;
\subfigure[Fig. 7: Achievable rate comparisons of considered algorithms versus ${b}$.] {
\label{mage04}
\includegraphics[width=0.43\columnwidth]{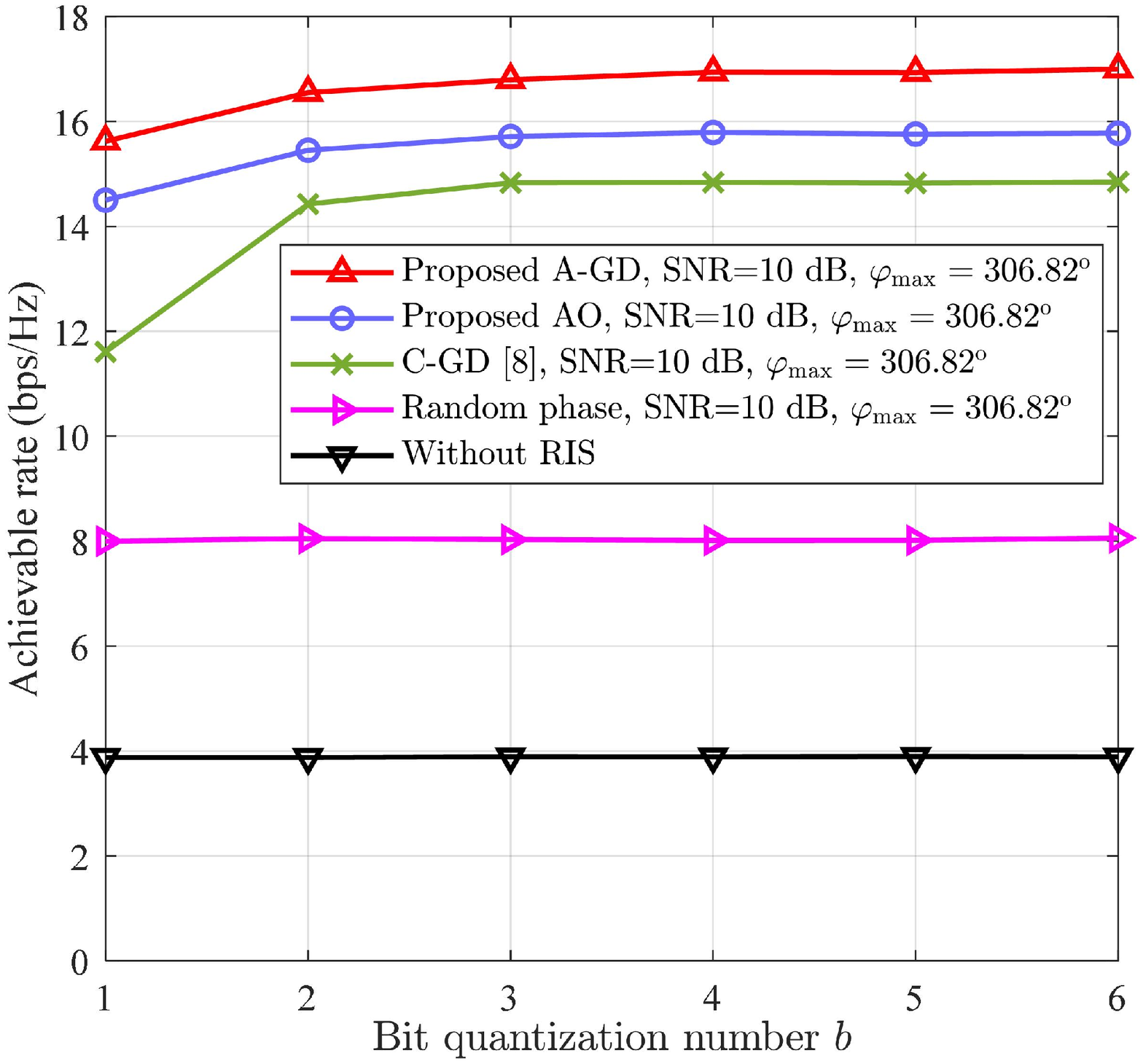}
}
\end{figure}


\makeatletter
\renewcommand{\@thesubfigure}{\hskip\subfiglabelskip}
\makeatother
\begin{figure} \centering
\subfigure[Fig. 8: Achievable rate comparisons of considered algorithms versus ${{N_{{\rm{MS}}}}}$.] {
 \label{mage05}
\includegraphics[width=0.43\columnwidth]{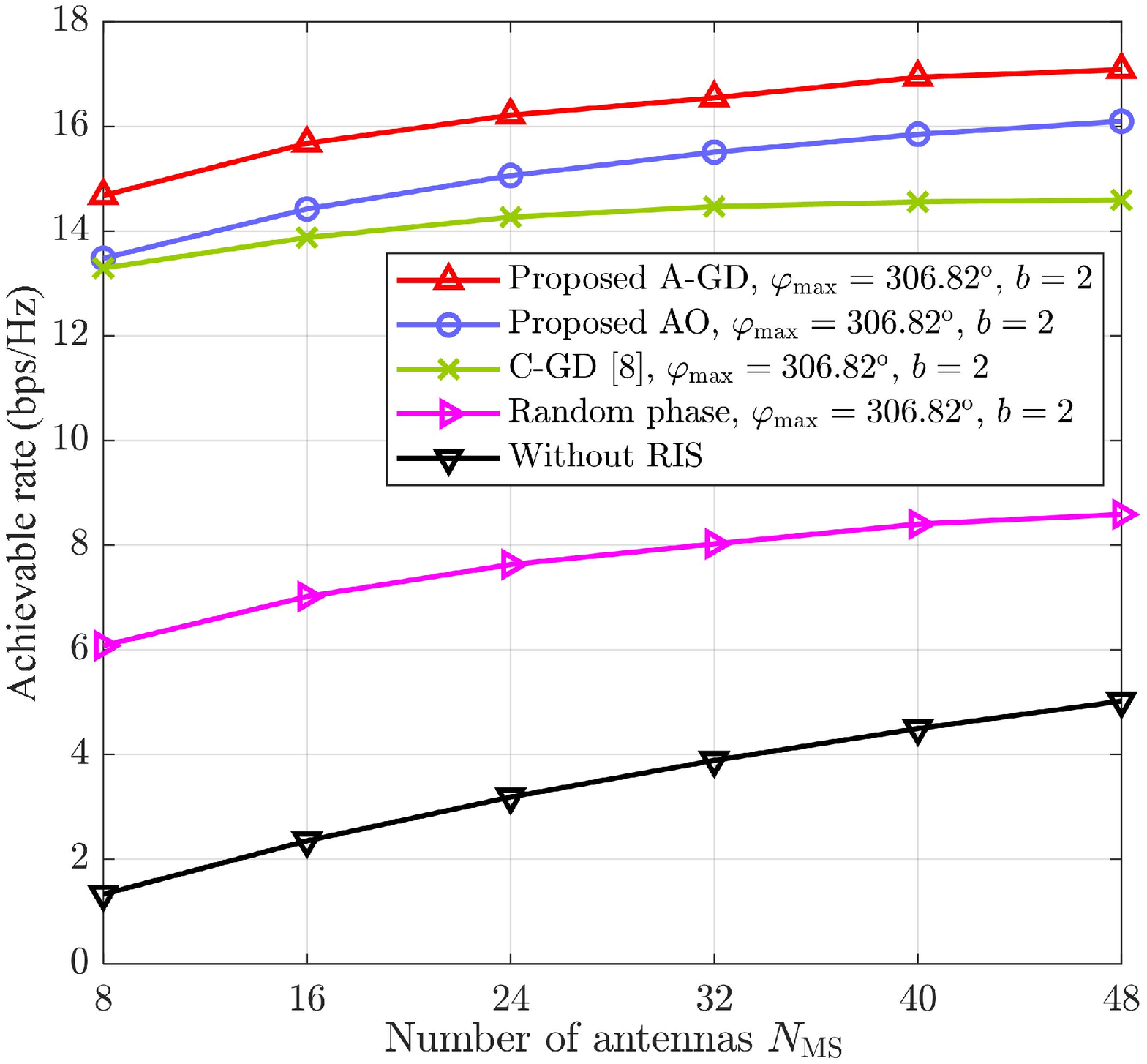}
} \;\;\;\;\;\;\;\;\;\;\;\;
\subfigure[Fig. 9: Achievable rate comparisons of considered algorithms versus ${{N_{{\rm{RIS}}}}}$.] {
\label{mage06}
\includegraphics[width=0.43\columnwidth]{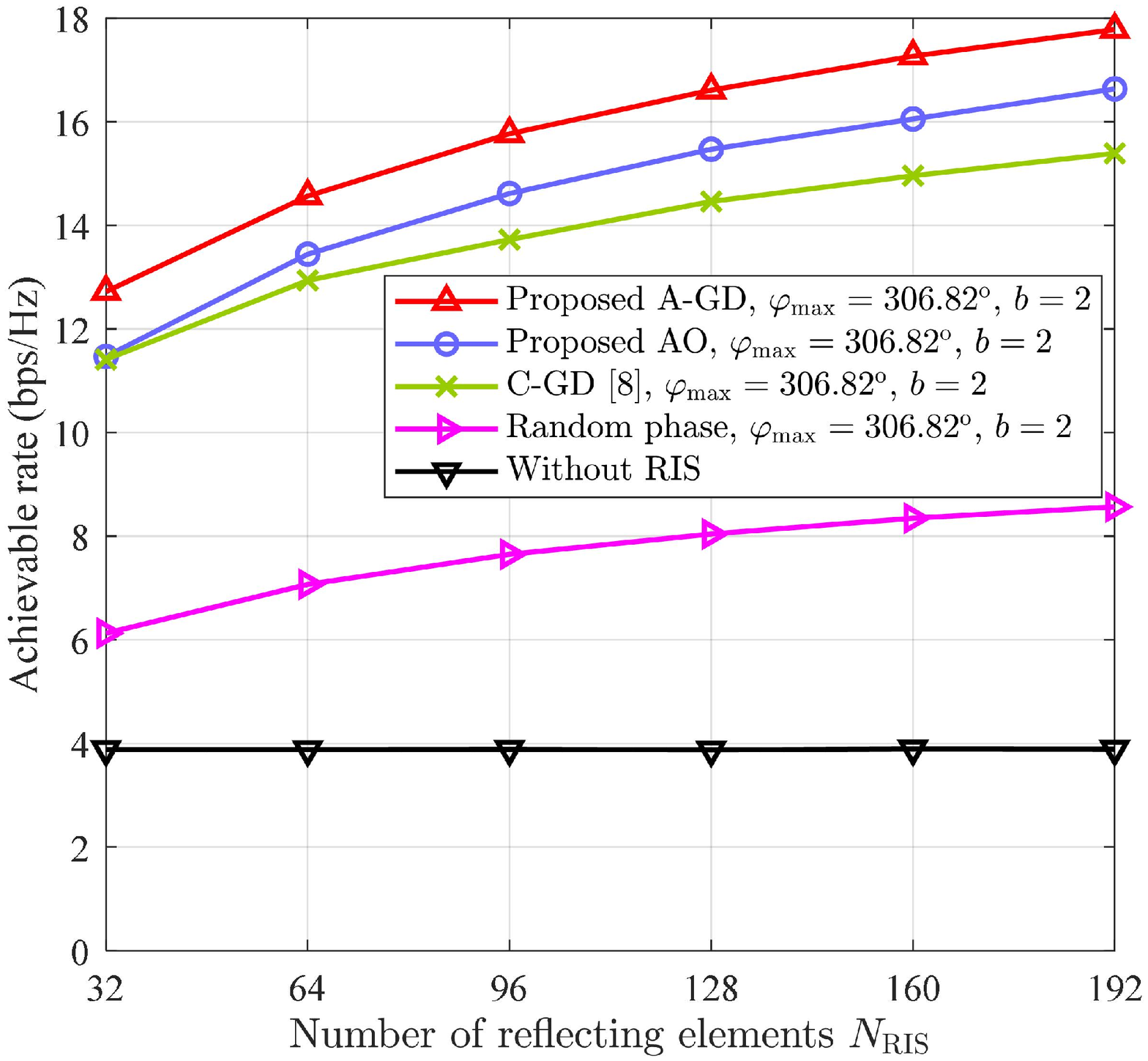}
}
\end{figure}


Fig. 8 displays the achievable rates of the considered schemes versus the different values of ${{N_{{\rm{MS}}}}}$ under the parameter settings of ${{N_{{\rm{BS}}}}{\rm{ = 512}}}$ and ${{N_{{\rm{RIS}}}}{\rm{ = 128}}}$. From Fig. 8, we can observe that the achievable rates of all the considered schemes become better when the number of antennas at the MS increases. Especially for the conventional THz system without RIS, the achievable rate can be improved from 1.33 bps/Hz to 5.02 bps/Hz once the number of antennas increases from ${{N_{{\rm{MS}}}}{\rm{ = 8}}}$ to ${{N_{{\rm{MS}}}}{\rm{ = 48}}}$. In addition, the achievable rate performance of our proposed algorithms (e.g., A-GD algorithm, AO algorithm) substantially outperform the benchmark schemes (e.g., C-GD algorithm, random phase scheme, the system without RIS) in case of different number of antennas. It is worth noting that the deployment of the number of antennas influences the system cost and power consumption, and thus selecting the appropriate number of antennas for different THz systems is still a significant issue.

Fig. 9 discusses the achievable rate comparisons of the considered schemes versus the number of RIS elements for the THz system configured as ${{N_{{\rm{BS}}}}{\rm{ = 512}}}$ and ${{N_{{\rm{MS}}}}{\rm{ = 32}}}$. From Fig. 9, we can note that the achievable rate of the THz system without RIS case has the worst performance due to the lack of reflecting signals, and remains unchanged along with the diverse values of ${{N_{{\rm{RIS}}}}}$. In terms of these RIS-enabled schemes, the achievable rates of our developed A-GD and AO algorithms outstrip the C-GD algorithm and the random phase scheme. Moreover, when the number of reflecting elements increases, the performance gaps between our proposed algorithms and the random phase scheme become much larger. Under the condition of ${{N_{{\rm{RIS}}}}{\rm{ = 192}}}$, our developed A-GD and AO algorithms realize around 9.21 bps/Hz and 8.06 bps/Hz performance enhancement compared with the random phase scheme. Hence, except for the performance gain brought by the RIS, our proposed optimization algorithms are capable of further improving the achievable rate of the RIS enabled THz MIMO system. Last but not least, it should be pointed out that the achievable rates of RIS-enabled algorithms will converge to finite values even if the number of reflecting elements goes to infinity, which is caused by the power budget limitation existing in practical communications.

\makeatletter
\renewcommand{\@thesubfigure}{\hskip\subfiglabelskip}
\makeatother
\begin{figure} \centering
\subfigure[Fig. 10: Convergence comparisons of considered algorithms versus the number of iterations.] {
 \label{mage07}
\includegraphics[width=0.43\columnwidth]{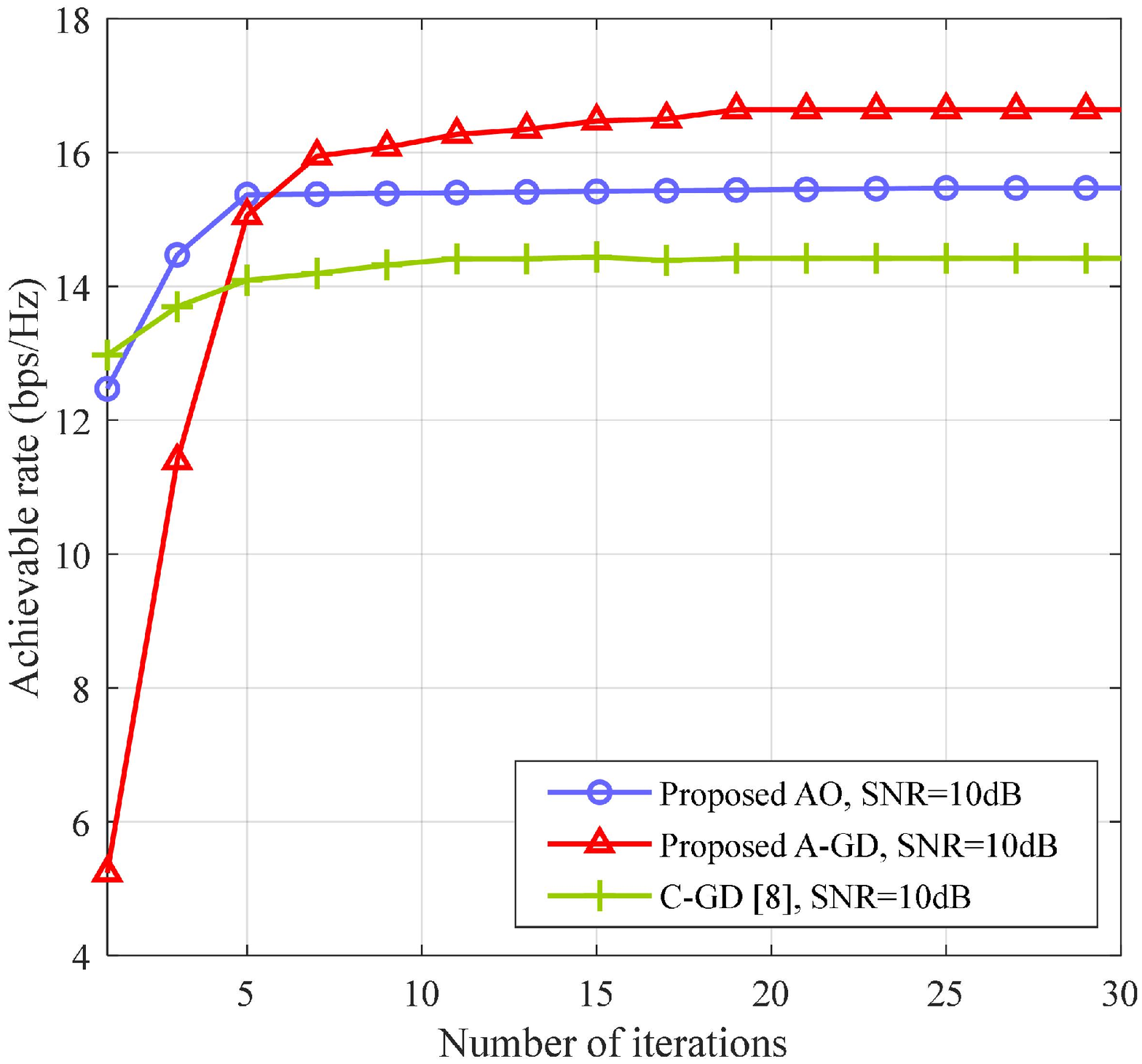}
} \;\;\;\;\;\;\;\;\;\;\;\;
\subfigure[Fig. 11: Complexity comparisons of considered algorithms versus ${{N_{{\rm{RIS}}}}}$.] {
\label{mage08}
\includegraphics[width=0.43\columnwidth]{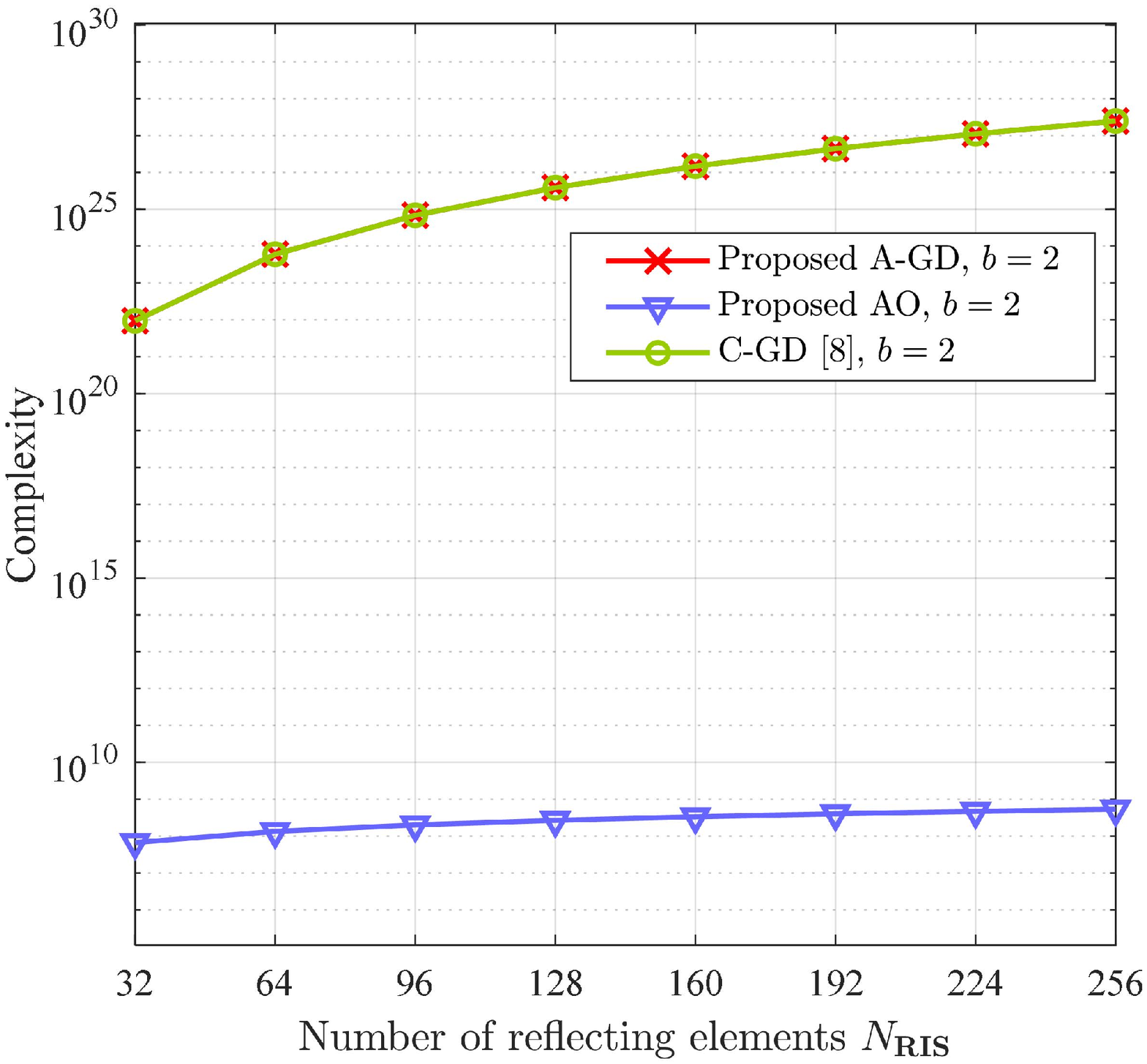}
}
\end{figure}

%
%


Fig. 10 depicts the convergence behavior of A-GD, C-GD and AO algorithms for the THz MIMO system with ${{N_{{\rm{BS}}}}{\rm{ = 512}}}$ and ${{N_{{\rm{MS}}}}{\rm{ = 32}}}$. From Fig. 10, we can know that the converged iterative numbers for A-GD, C-GD and AO algorithms are around ${{I_a} = 15}$, ${{I_c} = 12}$ and ${{I_o} = 3}$, respectively. Moreover, our proposed AO algorithm owns the best convergence performance, while its achievable rate outperforms the C-GD algorithm. When all the considered algorithms approach the converged points, the C-GD algorithm possesses the lowest achievable rate. On the basis of the convergence behavior, the complexity of the hybrid beamforming design algorithms along with the number of reflecting elements is presented in Fig. 11, where the vertical coordinate is expressed in the logarithmic coordinate. As shown in Fig. 11, the proposed AO algorithm has the lowest complexity. Instead, the A-GD algorithm and the C-GD algorithm suffer from abundant calculation burden. In addition, when ${{N_{{\rm{RIS}}}}}$ increases, the complexity of AO algorithm changes slowly while the A-GD algorithm and the C-GD algorithm grow rapidly. In the case of $ {N_{{\rm{RIS}}}} = 128$, the complexities of the AO, C-GD and A-GD are around $2.68 \times {10^8}$, $2.87 \times {10^{25}}$ and $2.87 \times {10^{25}}$, respectively. By observing Fig. 5 and Fig. 11, we can see that the AO algorithm realizes a better tradeoff between the achievable rate performance and computational complexity.

\section{Conclusion}
This paper jointly considered the hardware design and the hybrid beamforming optimization for the RIS-empowered THz MIMO communications. Considering the working principle of the RIS, we primarily probed into the characteristics of the practical graphene-based RIS, including the phase response and the reflecting amplitude. In light of these practical hardware constraints, the A-GD algorithm was developed to settle the hybrid beamforming design problem, and obtained much better performance than the C-GD algorithm. Furthermore, in order to depress the computational complexity, an AO algorithm was then raised by alternately performing the CBC algorithm and the linear search algorithm, which gained a better tradeoff between complexity and achievable rate performance. In the near future, our research work will concentrate on the practical measurements for the RIS empowered THz MIMO communications.


\begin{thebibliography}{1}
\bibitem{introduction_01}
K. B. Letaief, W. Chen, Y. Shi, et al., ``The Roadmap to 6G: AI Empowered Wireless Networks,'' \emph{IEEE Commun. Mag.}, vol. 57, no. 8, pp. 84-90, Aug. 2019.

\bibitem{introduction_02}
X. You et al., ``Towards 6G wireless communication networks: vision, enabling technologies, and new paradigm shifts,'' \emph{Sci. China Inf. Sci.}, vol. 64, no. 1, Jan. 2021.

\bibitem{introduction_03}
Z. Chen, X. Y. Ma, B. Zhang, et al., ``A survey on terahertz communications,'' \emph{China Commun.}, vol. 16, no. 2, pp. 1-35, Feb. 2019.

\bibitem{introduction_04}
I. F. Akyildiz, J. M. Jornet, and C. Han, ``Terahertz band: Next frontier for wireless communications,'' \emph{Phys. Commun. (Elsevier)}, vol. 12, no. 4, pp. 16-32, 2014.

\bibitem{introduction_06}
R. Piesiewicz et al., ``Short-Range Ultra-Broadband Terahertz Communications: Concepts and Perspectives,'' \emph{IEEE Antennas Propag. Mag.}, vol. 49, no. 6, pp. 24-39, Dec. 2007.


\bibitem{introduction_08}
I. F. Akyildiz, C. Han and S. Nie, ``Combating the Distance Problem in the Millimeter Wave and Terahertz Frequency Bands,'' \emph{IEEE Commun. Mag.}, vol. 56, no. 6, pp. 102-108, June. 2018.

\bibitem{introduction_09}
C. Han and Y. Chen, ``Propagation Modeling for Wireless Communications in the Terahertz Band,'' \emph{IEEE Commun. Mag.}, vol. 56, no. 6, pp. 96-101, Jun. 2018.

\bibitem{yanlongfei_01}
C. Huang, A. Zappone, G. C. Alexandropoulos, et al., ``Reconfigurable Intelligent Surfaces for Energy Efficiency in Wireless Communication,'' \emph{IEEE Trans. Wirel. Commun.}, vol. 18, no. 8, pp. 4157-4170, Aug. 2019.

\bibitem{introduction_10}
T. Cui, M. Qi, X. Wan, et al., ``Coding metamaterials, digital metamaterials and programmable metamaterials,'' \emph{Light Sci. Appl.}, vol. 3, pp. e218, 2014.

\bibitem{introduction_10-1}
C. Huang, Z. Yang, G. C. Alexandropoulos, et al., ``Multi-hop RIS-Empowered Terahertz Communications: A DRL-based Hybrid Beamforming Design,'' \emph{IEEE J. Sel. Areas Commun.}, vol. 39, no. 6, pp. 1663-1677, June 2021.

\bibitem{introduction_12}
C. Liaskos, S. Nie, A. Tsioliaridou, A. Pitsillides, S. Ioannidis and I. Akyildiz, ``A New Wireless Communication Paradigm through SoftwareControlled Metasurfaces,'' \emph{IEEE Commun. Mag.}, vol. 56, no. 9, pp. 162-169, Sept. 2018.

\bibitem{introduction_13}
L. Dai, B. Wang, M. Wang, et al., ``Reconfigurable Intelligent Surface-Based Wireless Communications: Antenna Design, Prototyping, and Experimental Results,'' \emph{IEEE Access}, vol. 8, pp. 45913-45923, 2020.

\bibitem{introduction_13-1}
C. Huang, S. Hu, G. C. Alexandropoulos, et al., ``Holographic MIMO surfaces for 6G wireless networks: opportunities, challenges, and trends,'' \emph{IEEE Wireless Communications}, vol. 27, no. 5, pp. 118-125, Oct. 2020.


\bibitem{introduction_14}
G. Wang, F. Gao, R. Fan and C. Tellambura, ``Ambient Backscatter Communication Systems: Detection and Performance Analysis,'' \emph{IEEE Trans. Commun.}, vol. 64, no. 11, pp. 4836-4846, Nov. 2016.


\bibitem{introduction_16}
C. Tienda, M. Arrebola, J. A. Encinar and G. Toso, ``Analysis of a dualreflect array antenna,'' \emph{IET Microwaves, Antennas and Propagation}, vol. 5, no. 13, pp. 1636-1645, Oct. 2011.

\bibitem{introduction_17}
F. Gao, T. Cui and A. Nallanathan, ``On channel estimation and optimal training design for amplify and forward relay networks,'' \emph{IEEE Trans. Wirel. Commun.}, vol. 7, no. 5, pp. 1907-1916, May 2008.

\bibitem{introduction_21}
X. Ma, Z. Chen, W. Chen, et al., ``Joint Channel Estimation and Data Rate Maximization for Intelligent Reflecting Surface Assisted Terahertz MIMO Communication Systems,'' \emph{IEEE Access}, vol. 8, pp. 99565-99581, May 2020.

\bibitem{introduction_21-1}
Z. He and X. Yuan, ``Cascaded Channel Estimation for Large Intelligent Metasurface Assisted Massive MIMO,''  \emph{IEEE Wireless Commun. Lett.}, vol. 9, no. 2, pp. 210-214, Feb. 2020.

\bibitem{introduction_21-2}
H. Liu, X. Yuan and Y. -J. A. Zhang, ``Matrix-Calibration-Based Cascaded Channel Estimation for Reconfigurable Intelligent Surface Assisted Multiuser MIMO,'' \emph{IEEE J. Sel. Areas Commun.}, vol. 38, no. 11, pp. 2621-2636, Nov. 2020.

\bibitem{introduction_21-3}
C. Huang, G. C. Alexandropoulos, A. Zappone, M. Debbah and C. Yuen, ``Energy Efficient Multi-User MISO Communication Using Low Resolution Large Intelligent Surfaces,'' in \emph{Proc. IEEE GLOBECOM
Workshops}, pp. 1-6, 2018.

\bibitem{introduction_22}
Q. Wu, R. Zhang,  ``Intelligent reflecting surface enhanced wireless network: Joint active and passive beamforming design,'' in \emph{Proc. IEEE GLOBECOM}, pp 1-6, 2018.

\bibitem{introduction_23}
X. Ma, Z. Chen, W. Chen, et al., ``Intelligent Reflecting Surface Enhanced Indoor Terahertz Communication Systems,'' \emph{Nano Commun. Netw.}, vol. 24, pp. 100284, May. 2020.

\bibitem{introduction_24}
C. Pan et al., ``Multicell MIMO Communications Relying on Intelligent Reflecting Surfaces,'' \emph{IEEE Trans. Wirel. Commun.}, vol. 19, no. 8, pp. 5218-5233, Aug. 2020

\bibitem{introduction_29}
H. Shen, W. Xu, S. Gong, Z. He and C. Zhao, ``Secrecy Rate Maximization for Intelligent Reflecting Surface Assisted Multi-Antenna Communications,'' \emph{IEEE Commun. Lett.}, vol. 23, no. 9, pp. 1488-1492, Sept. 2019.

\bibitem{introduction_29-1}
L. Liang, M. Qi, J. Yang, et al., ``Anomalous Terahertz Reflection and Scattering by Flexible and Conformal Coding Metamaterials,'' \emph{Adv. Opt. Mater.}, vol. 3, no. 10, pp. 1311-1311, 2015.

\bibitem{introduction_29-2}
S. Liu, L. Zhang, Q. Yang, et al., ``Frequency dependent dual functional coding metasurfaces at terahertz frequencies,'' \emph{Adv. Opt. Mater.}, vol. 4, no. 12, pp. 1965-1973, 2016.

%

\bibitem{introduction_31}
P. Wang, J. Fang, L. Dai and H. Li, ``Joint Transceiver and Large Intelligent Surface Design for Massive MIMO MmWave Systems,'' \emph{IEEE Trans. Wirel. Commun.}, vol. 20, no. 2, pp. 1052-1064, Feb. 2021.

\bibitem{introduction_31-1}
Q. Wu and R. Zhang, ``Beamforming Optimization for Wireless Network Aided by Intelligent Reflecting Surface With Discrete Phase Shifts,'' \emph{IEEE Trans. Commun.}, vol. 68, no. 3, pp. 1838-1851, Mar. 2020.

\bibitem{introduction_31-2}
C. Huang, R. Mo and C. Yuen, ``Reconfigurable Intelligent Surface Assisted Multiuser MISO Systems Exploiting Deep Reinforcement Learning,'' \emph{IEEE J. Sel. Areas Commun.}, vol. 38, no. 8, pp. 1839-1850, Aug. 2020.

\bibitem{introduction_32}
A. Goldsmith, S. A. Jafar, N. Jindal and S. Vishwanath, ``Capacity limits of MIMO channels,'' \emph{IEEE J. Sel. Areas Commun.}, vol. 21, no. 5, pp. 684-702, Jun. 2003.

\bibitem{introduction_33}
O. E. Ayach, S. Rajagopal, S. Abu-Surra, et al., ``Spatially Sparse Precoding in Millimeter Wave MIMO Systems,'' \emph{IEEE Trans. Wirel. Commun.}, vol. 13, no. 3, pp. 1499-1513, Mar. 2014.


\bibitem{system_01}
A. A. M. Saleh and R. Valenzuela, ``A Statistical Model for Indoor Multipath Propagation,'' \emph{IEEE J. Sel. Areas Commun.}, vol. 5, no. 2, pp. 128-137, Feb. 1987.

\bibitem{system_02}
C. Han, A. O. Bicen, I. F. Akyildiz, ``Multi-Ray Channel Modeling and Wideband Characterization for Wireless Communications in the Terahertz Band,'' \emph{IEEE Trans. Wirel. Commun.}, vol. 14, no. 5, pp. 2402-2412, Dec. 2015.

\bibitem{system_02-1}
W. Tang, M. Chen, X. Chen, et al., ``Wireless Communications With Reconfigurable Intelligent Surface: Path Loss Modeling and Experimental Measurement,'' \emph{IEEE Trans. Wirel. Commun.}, vol. 20, no. 1, pp. 421-439, Jan. 2021.

\bibitem{system03}
Z. Wang, L. Liu and S. Cui, ``Channel Estimation for Intelligent Reflecting Surface Assisted Multiuser Communications: Framework, Algorithms, and Analysis,'' \emph{IEEE Trans. Wirel. Commun.}, vol. 19, no. 10, pp. 6607-6620, Oct. 2020.


\bibitem{yaojia_01}
H. Chen, A. J. Taylor, N. Yu, ``A review of metasurfaces: physics and applications,'' \emph{Rep. Prog. Phys.}, vol. 79, no. 7, pp. 076401, Jun. 2016.



\bibitem{yaojia_08}
H. Wong, D. Akinwande, ``Carbon Nanotube and Graphene Device Physics,'' \emph{Cambridge University Press}, 2010.

\bibitem{yaojia_09}
D. K. Efetov, P. Kim, ``Controlling electron-phonon interactions in graphene at ultrahigh carrier densities,'' \emph{Phys. Rev. Lett.}, vol. 105, no. 25, pp. 256805, Dec. 2010.

\bibitem{yaojia_10}
S. Lee, M. Choi, T. Kim et al., ``Switching teraherz waves with gate-controlled active graphene metamaterials,'' \emph{Nat. Mater}, vol. 11, pp. 936-941, Sep. 2012.

\bibitem{yaojia_11}
Y. Yang, W. Wang, P. Moitra et al., ``Dielectric Meta-Reflectarray for Broadband Linear Polarization Conversion and Optical Vortex Generation,'' \emph{Nano Lett.}, vol.14, no.3, pp. 1394-1399, Feb. 2014.

\bibitem{yaojia_12}
A. Pors, S. I. Bozhevolnyi, ``Plasmonic metasurfaces for efficient phase control in reflection,'' \emph{Opt. Express}, vol. 21, no. 22, pp. 27438-27451, Nov. 2013.

\bibitem{yaojia_13}
Z. Li, K. Yao, F. Xia et al., ``Graphene Plasmonic Metasurfaces to Steer Infrared Light,'' \emph{Sci. Rep.}, vol. 5, no. 12423, Jul. 2015.

\bibitem{yaojia_14}
B. Di, H. Zhang, L. Song, et al., ``Hybrid Beamforming for Reconfigurable Intelligent Surface based Multi-User Communications: Achievable Rates With Limited Discrete Phase Shifts,'' \emph{IEEE J. Sel. Areas Commun.}, vol. 38, no. 8, pp. 1809-1822, Aug. 2020.


\bibitem{yanlongfei_02}
X. Yu, J. Shen, J. Zhang, and K. B. Letaief, ``Alternating Minimization Algorithms for Hybrid Precoding in Millimeter Wave MIMO Systems,'' \emph{IEEE J. Sel. Top. Signal Process.}, vol. 10, no. 3, pp. 485-500, Feb. 2016.

\bibitem{yanlongfei_03}
L. Yan, C. Han and J. Yuan, ``A Dynamic Array-of-Subarrays Architecture and Hybrid Precoding Algorithms for Terahertz Wireless Communications,'' \emph{IEEE J. Sel. Areas Commun.}, vol. 38, no. 9, pp. 2041-2056, Sept. 2020.

\bibitem{yanlongfei_04}
X. Zhang, A. F. Molisch, and S.-Y. Kung, ``Variable-phase-shift-based RF-baseband codesign for MIMO antenna selection,'' \emph{IEEE Trans. Signal Process.}, vol. 53, no. 11, pp. 4091-4103, Nov. 2005.

\bibitem{yanlongfei_05}
C. Rusu, R. Mndez-Rial, N. Gonzlez-Prelcic, and R. W. Heath, ``Low Complexity Hybrid Precoding Strategies for Millimeter Wave Communication Systems,'' \emph{IEEE Trans. Wirel. Commun.}, vol. 15, no. 12, pp. 8380-8393, Dec. 2016.

\bibitem{yanlongfei_06}
C. Lin and G. Y. Li, ``Adaptive Beamforming With Resource Allocation for Distance-Aware Multi-User Indoor Terahertz Communications,'' \emph{IEEE Trans. Commun.}, vol. 63, no. 8, pp. 2985-2995, Aug. 2015.

\end{thebibliography}
\end{document}